\documentclass[a4paper,onecolumn,preprintnumbers,superscriptaddress,nofootinbib,aps,prd,floatfix,10pt]{revtex4}

\usepackage{graphicx,footmisc,hyperref,subfigure}
\usepackage{amsmath,amssymb,mathrsfs,slashed,braket,dsfont}
\usepackage{multirow,booktabs}
\usepackage{feynmp}
\newcommand\numberthis{\addtocounter{equation}{1}\tag{\theequation}}
\usepackage{gensymb}
\allowdisplaybreaks
\usepackage{multirow}
\usepackage{tabularx}

\makeatletter
\@ifundefined{pdfoutput}{}{\DeclareGraphicsRule{*}{mps}{*}{}}
\def\l@subsubsection#1#2{}
\makeatother

\begin{document}
\title{SMEFT Corrections to $Z$ Boson Decays}

\begin{abstract}
We compute the one-loop corrections to $Z$ decay properties from dimension-6 operators in the Standard Model
Effective Field Theory (SMEFT) that
contribute also to anomalous 3-gauge boson couplings and examine the relative sensitivity of the two processes
to the anomalous couplings.  The size of the contributions is of order  a few percent, of the same size as Standard Model electroweak corrections.  
 This is part of a program of computing electroweak quantities to one-loop
in the SMEFT:  these calculations are needed for a future  global fit to limit the coefficients of the dimension-six Wilson coefficients consistently
at one loop. \end{abstract}

\author{Sally Dawson}
\email{dawson@bnl.gov}
\affiliation{Department of Physics, Brookhaven National Laboratory, Upton, N.Y., 11973\\[0.1cm]}
\author{Ahmed Ismail} \email{aismail@pitt.edu}
\affiliation{PITT-PACC, University of Pittsburgh, Pittsburgh, PA., 15260\\[0.1cm]}
\preprint{PITT-PACC-1813}

\maketitle
\newpage
\tableofcontents
\newpage

\section{Introduction}
\label{sec:intro}

The development of the precision electroweak program at the LHC is a major task for the coming decade.  At present, the interactions of the Higgs boson and the electroweak gauge bosons appear to have approximately Standard Model (SM)  like interactions and there is no sign of new massive particles. These points together imply  that deviations from the SM can be analyzed in an effective field theory 
framework\cite{Giudice:2007fh,Brivio:2017vri}.

In the  Standard Model Effective Field Theory (SMEFT), deviations from the SM are parameterized in terms of a tower of higher dimension operators, $O_k^d$,
\begin{equation}
\mathcal{L}=\mathcal{L}_{SM}+\Sigma_d\Sigma_k{C_k^d O_k^d\over \Lambda^{d-4}}\, ,
\end{equation}
where the operators, $O_k^d$, contain only SM fields and are invariant under $SU(3)\times SU(2)\times U(1)$.  
The complete set of dimension-$6$ operators was first compiled in Refs. \cite{Buchmuller:1985jz,Grzadkowski:2010es} and the Feynman rules in this
basis ("Warsaw Basis") are conveniently given in Ref. \cite{Dedes:2017zog}.  The new physics is completely contained in the coefficient
functions, $C_k^d$.
The scale of the assumed UV complete theory is $\Lambda$, and we assume $\Lambda \gg v=246$~GeV.  
For a weakly coupled theory, the corrections to SM  predictions are dominated by the dimension-$6$ contributions.

Predictions for Higgs production and decay, along with $VV$ ($W^\pm,Z,\gamma$) interactions are well known at tree level 
in the SMEFT\cite{Giudice:2007fh,Brivio:2017vri,Falkowski:2015fla,Contino:2014aaa}.
Including also contributions to the oblique parameters, limits on the allowed sizes of the SMEFT coefficients can be extracted
in a global fit to Higgs signal rates and gauge boson pair production\cite{Butter:2016cvz,Ellis:2018gqa,Berthier:2016tkq,Falkowski:2017pss}. A precision Higgs and electroweak physics program, however, requires SMEFT calculations beyond the leading order if matching between the experimental results and theory is to be eventually done at the few percent level.  

The program of calculating SMEFT quantities beyond leading order is in its infancy.  One-loop calculations exist for $H\rightarrow \gamma
\gamma$\cite{Hartmann:2015aia,Hartmann:2015oia,Dedes:2018seb}, $H\rightarrow b {\overline b}$\cite{Gauld:2016kuu,Gauld:2015lmb}
 and the 
unphysical $H\rightarrow ZZ$ and $H\rightarrow W^+W^-$ processes\cite{Dawson:2018pyl,Dawson:2018liq}.  The  one-loop Yukawa, $y_t$, and $\lambda={M_H^2\over 2 v^2}$ contributions to  $Z$ decays are also known\cite{Hartmann:2016pil}. 
In addition to effects in the electroweak sector, one-loop contributions from top-quark operators can significantly affect Higgs production rates at the LHC\cite{Degrande:2012gr,Vryonidou:2018eyv}. 
 
 In this paper, we compute the $1$-loop corrections to 
the partial $Z$ decay widths due to the dimension-$6$ operators that contribute to $pp\rightarrow W^+W^-$ and compare the sensitivity of
the two processes. These operators are particularly interesting because for transverse gauge boson production they contribute to different helicity amplitudes~\cite{Azatov:2016sqh,Baglio:2017bfe}, such that their interference with the SM does not grow with energy unless decays or higher order corrections are considered~\cite{Azatov:2017kzw,Panico:2017frx}.
 Along with anomalous 3-gauge boson couplings, we include in our calculation the shifts in the $Z$ decay widths due to anomalous fermion couplings, which have important
contributions not only to the $Z$ widths\cite{Dawson:1994fa}, but also to gauge boson pair production\cite{Baglio:2017bfe,Zhang:2016zsp,Alves:2018nof}.   
Low energy data places strong limits on deviations from the SM and information from $Z$ decays is particularly interesting due to the precision
of the LEP measurements.   Consistent fits to the LEP data  require the inclusion of the complete set of SMEFT operators, along with the 
one-loop predictions.  Our calculation is a step in this direction, and is related to previous studies of the loop effects of gauge boson self-couplings on precision electroweak observables~\cite{Hagiwara:1992eh,Hagiwara:1993ck,Alam:1997nk,Mebane:2013cra,Mebane:2013zga}. 

In Section \ref{sec:smeft1}, we review the basics of the one-loop SMEFT calculation and in Section \ref{sec:zdec} the calculation of
$Z\rightarrow f {\overline f}$ in the SMEFT is summarized, with analytic formulae presented in a series of appendices.  Numerical results are given in Section \ref{sec:zdec}.
 
\section{SMEFT at one-loop}
\label{sec:smeft1}

In this work we consider modifications of the  $Z f {\overline f}$  and $W^+W^-V$ ($V=Z,\gamma$) vertices.  We consider only operators that
contribute  to both $q {\overline q}\rightarrow W^+W^-$ \cite{Baglio:2017bfe,Alves:2018nof} and  to $Z\rightarrow f {\overline f}$.

The fermion vertices can be
parameterized as,
\begin{eqnarray}
  \mathcal{L}_f&=&g_ZZ_\mu\biggl[
  \biggl(g_L^{Zf}+\delta g_{L}^{Zf}\biggr)
  {\overline f}_L\gamma_\mu f_L\
 +\biggl[g_R^{Zf}+\delta g_{R}^{Zf}\biggr]
  {\overline f}_R\gamma_\mu f_R +(f\rightarrow f^\prime)\biggr]\nonumber \\
  &&+{g\over \sqrt{2}}\biggl\{W_\mu\biggl[(1+\delta g_{L}^W){\overline f}_L\gamma_\mu f_L^\prime
  +\delta g_R^W
  {\overline f}_R\gamma_\mu f_R^\prime\biggr] +h.c.\biggr\}\, ,
  \label{eq:dgdef}
  \end{eqnarray}
 where $g_Z\equiv e/(c_W^{}s_W^{})= g/c_W$ and $f$  ($f^\prime$) denotes up-type (down-type) quarks.
 The SM fermion couplings are:
\begin{eqnarray}
g_R^{Zf}&=&-s_W^2 Q_f\quad{\rm and}\quad g_L^{Zf}=T_3^f -s_W^2 Q_f,
\end{eqnarray}
where $T_3^f=\pm \displaystyle \frac{1}{2}$ and  $Q_f$  are the weak isospin and electric
 charge of the fermions, respectively.

Assuming CP conservation, the most general Lorentz invariant $3-$gauge
boson couplings can be written
as~\cite{Gaemers:1978hg,Hagiwara:1986vm}
\begin{eqnarray}
 L_{V}=
-ig_{WWV}\left(g_1^V\left(W^+_{\mu\nu}W^{-\mu}V^\nu-W_{\mu\nu}^-W^{+\mu}V^\nu\right)+\kappa^VW^+_\mu
            W^-_\nu V^{\mu\nu}+\frac{\lambda^V}{M^2_W}W^+_{\rho\mu}{W^{-\mu}}_\nu V^{\nu\rho}\right),
\label{eq:lagdef}
\end{eqnarray}  
where $g_{WW\gamma}=e$ and $g_{WWZ}=g c_W$.
For the $3-$gauge boson
couplings we define $g_1^V = 1+\delta g_1^V$, $\kappa_{}^V=
1+\delta\kappa_{}^V$, and in the SM, $\delta g_1^V = \delta\kappa_{}^V
= \lambda_{}^V = 0$. Because of gauge invariance we always have
$\delta g_1^\gamma = 0$. We assume $SU(2)$ invariance, which implies that
the coefficients are related by,
\begin{eqnarray}
\delta g_L^W&=&\delta g_L^{Zf}-\delta g_L^{Zf'},
\nonumber \\
\delta g_1^Z&=& \delta \kappa_{}^Z+{s_W^2\over c_W^2}\delta \kappa_{}^\gamma,
\nonumber \\
\lambda_{}^\gamma &=& \lambda_{}^Z,
\label{eq:su2rel}
\end{eqnarray}
leaving three independent effective couplings.

We work in the Warsaw
basis~\cite{Buchmuller:1985jz,Grzadkowski:2010es} and the dimension-$6$ operators
contributing to the 3-gauge boson vertices are, 
\begin{eqnarray}
\mathcal{O}_{W}&=& \epsilon^{abc} W_\mu^{a\nu}W_\nu^{b\rho}W_\rho^{c\mu}\nonumber \\
\mathcal{O}_{HWB}&=& \Phi^\dagger\sigma^a\Phi W^a_{\mu\nu}B^{\mu\nu}\nonumber \\
\label{eq:ops}
\end{eqnarray}
where  $D_\mu \Phi=(\partial_\mu -i\,\frac{g}{2}\sigma^a
W^a_\mu-i\frac{g'}{2}B_\mu)\Phi$, $W^a_{\mu\nu}=\partial_\mu W^a_\nu
-\partial_\nu W^a_\mu+g\varepsilon^{abc}W^b_\mu W^c_\nu$,  and $\Phi$  is the
Higgs doublet field with a vacuum expectation value $\langle\Phi\rangle =
(0,v/\sqrt{2})^{\rm T}$.  
Two other operators involving the Higgs and gauge bosons make important contributions to the 
effective $Zf\bar{f}$ vertices, 
\begin{eqnarray}
\mathcal{O}_{HW}&=& (\Phi^\dagger\Phi)W^a_{\mu\nu}W^{a\mu\nu}\nonumber \\
\mathcal{O}_{HB}&=&(\Phi^\dagger\Phi)B_{\mu\nu}B^{\mu\nu}\, .
\label{ops:ex}
\end{eqnarray}
 $\mathcal{O}_{HW}$ and $\mathcal{O}_{HB}$ contribute to the 1-loop renormalization of the input parameters, as discussed in the next section.

   We take as our input parameters $M_W, M_Z$ and $G_{\mu}$. All other parameters are defined in terms of the input parameters.
   The Lagrangian of interest to us is:
\begin{eqnarray}
\mathcal{L}&=&-\frac14 W_{\mu\nu}^a W_{\mu\nu}^a-\frac14 B_{\mu\nu}B_{\mu\nu}
+\frac1{\Lambda^2}\left(C_{HW} \mathcal{O}_{HW} +C_{H B} \mathcal{O}_{HB} +C_{HWB} \mathcal{O}_{HWB} 
+C_{W} \mathcal{O}_{W}
\right).
\end{eqnarray}
We define  ``barred'' fields, ${\overline W}_\mu\equiv (1-C_{H W} v^2/\Lambda^2)W_\mu$ and ${\overline B}_\mu\equiv (1-C_{H B}v^2/\Lambda^2)B_\mu$ and ``barred'' gauge couplings,  ${\overline g}\equiv (1+C_{H W} v^2/\Lambda^2)g$ and ${\overline g^\prime}\equiv (1+C_{H B}v^2/\Lambda^2)g^\prime$ so that ${\overline W}_\mu {\overline g}= W_\mu g$ and ${\overline B}_\mu {\overline g^\prime}= B_\mu g^\prime $. The ``barred'' fields have their kinetic terms properly normalized and the covariant derivatives have the canonical form.
The masses of the W and Z fields (poles of the propagators) are, in 
terms of the ``barred'' couplings \cite{Dedes:2017zog,Alonso:2013hga},
\begin{eqnarray}
M_W^2&=&\frac{{\overline g}^2 v^2}4,\nonumber\\
M_Z^2&=&\frac{({\overline g^\prime}^2+{\overline g}^2) v^2}4+\frac{v^4}{\Lambda^2}\left(\frac18 ({\overline g^\prime}^2+{\overline g}^2) C_{H D}
+\frac12 {\overline g^\prime}{\overline g}C_{H WB} \right).
\label{eq:wdef}
\end{eqnarray}
The extra terms in the definition of the $Z$ mass are due to the rotation, $(W_\mu^3,B_\mu)\to(Z_\mu,A_\mu)$,
 that is proportional to $C_{HWB}$\footnote{We will neglect the  contribution to $M_Z$  that is proportional to $C_{H D}$.}.
 We can define $\cos\theta_W\equiv c_W$ in terms of $M_W$ and $M_Z$,
 \begin{eqnarray}
 c_W^2&\equiv& {M_W^2\over M_Z^2}\nonumber \\
 &=&{{\overline g}_2^2\over 
 ({\overline g}_1^2+{\overline g}_2^2) }\biggl(1+{\delta c_W^2\over c_W^2}\biggr)
 \, ,
 \end{eqnarray}
and ${\delta s_W^2\over s_W^2}=-{\delta c_W^2\over c_W^2}$.
 Comparing with Eq. \ref{eq:wdef}, 
 \begin{equation}
{\delta s_W^2} =-{s_{W} c_{W}\over c_{W}^2-s_{W}^2}{v^2\over\Lambda^2}C_{HWB}\, .
\label{eq:sdef}
\end{equation}
In Eq. \ref{eq:sdef} we can   use, $c_{W}={M_W\over M_Z}$ to ${\cal {O}}({v^2\over \Lambda^2})$.

We
find  the following mappings between the SMEFT coefficients, $C_{HWB}$ and $C_W$,  and the effective couplings,
\begin{eqnarray}
\delta g_1^Z &=& -{\delta s_W^2\over c_W^2}\nonumber\\
\delta \kappa^Z&=& -2\delta s_W^2\nonumber \\
\delta \kappa_{}^\gamma &=&-{c_W^2-s_W^2\over s_W^2}\delta s_W^2  \nonumber \\ 
\lambda_{}^V&=& \frac{v}{\Lambda^2} 3 M_W^{} C_{W}\nonumber\\
\delta g_L^W &=&  \delta s_W^2\nonumber\\ 
\delta g_R^W&=&0\, \nonumber\\
\delta
  g^{Zf}_{L,R}&=&Q_f \delta s_W^2 ,\nonumber\\
\label{eq: mapcoef}
\end{eqnarray}
The shifts including the SMEFT operators that we have omitted can be found in Refs.~\cite{Berthier:2015oma,Zhang:2016zsp}.
\section{Results}
\label{sec:zdec}

At tree level, the decay amplitude for $Z \to f(p) \bar{f}(p')$ in the SMEFT is,  (including only those terms that contribute  also to 3- gauge boson  vertices), 
\begin{equation}
\mathcal{M}_0 =2M_{Z0}\sqrt{\sqrt{2}G_{\mu 0}}\biggl\{T_3^f-Q_f\biggl(1-{M_{W0}^2\over M_{Z0}^2}\biggr)
+Q_f {M_{W0}\over M_{Z0}}\sqrt{1-{M_{W0}^2\over M_{Z0}^2}}{v^2\over\Lambda^2}
C_{HWB} \biggr\} \bar{u}(p) \slashed{\epsilon}^*(p + p') v(p')\, ,
\end{equation}
where  the subscript $"0"$ indicates the unrenormalized tree level value, and in the $C_{HWB}$ term we can take $v^2={1\over \sqrt{2} G_\mu}$.

At one loop, there are contributions from corrections to the input parameters and fields to $\mathcal{M}_0$, $Z-\gamma$ mixing, and the  one-particle irreducible loop corrections to the decay, $\mathcal{M}_1$.  The virtual decay amplitude is, 
\begin{equation}
\begin{split}
\mathcal{M}_{\mathrm{1-loop}} &= \left( 1 + \delta C_{HWB} \frac{\partial}{\partial C_{HWB}} + \delta G_\mu \frac{\partial}{\partial G_\mu} + \delta M_Z^2 \frac{\partial}{\partial M_Z^2} + \delta M_W^2 \frac{\partial}{\partial M_W^2} + \frac{1}{2} \delta Z_Z + \delta Z_f \right) \mathcal{M}_0 \\&\quad - \mathcal{M}_\gamma \frac{\Pi_{\gamma Z}(M_Z^2)}{M_Z^2} + \mathcal{M}_1\, .
\label{eq:zdecayamp}
\end{split}
\end{equation}
Here $\mathcal{M}_\gamma$ is the amplitude for $\gamma \to f(p) \bar{f}(p')$, which is
\begin{equation}
\mathcal{M}_\gamma ={2M_W }\sqrt{\sqrt{2}G_\mu}Q_f\biggl\{\sqrt{1-{M_W^2\over M_Z^2}}-{M_W\over M_Z}{v^2\over\Lambda^2}C_{HWB}\biggr\} \bar{u}(p) \slashed{\epsilon}^*(p + p') v(p')\, ,
\end{equation}
in the SMEFT. 
In Eq. \ref{eq:zdecayamp}, $Z_f$ and $Z_Z$ are the wavefunction renormalizations of the external $Z$ boson and the fermions.  We use on-shell renormalization for
all quantities, except for the Wilson coefficients which are renormalized using ${\overline{MS}}$ subtraction. 
In general,  the coefficients are renormalized as\cite{Alonso:2013hga,Grojean:2013kd}, 
\begin{equation}
C_i(\mu)=C_{0,i}
-{1\over 32\pi^2  \hat{\epsilon}}\gamma_{ij}C_j,
\end{equation}
where $\mu$ is the renormalization scale, $\gamma_{ij}$ is the one-loop anomalous dimension and $\hat{\epsilon}^{-1}\equiv\epsilon^{-1}-\gamma_E+\log(4\pi)$ is related to the regulator $\epsilon$ for integrals evaluated in $d=4-2\epsilon$ dimensions.

The renormalization of $G_\mu$ in the SMEFT,
including both logarithms and constant contributions,
can be found in the appendix of Ref. \cite{Dawson:2018pyl}. The shifts in the SM input parameters as well as the external field wave function renormalizations follow from the 2-point functions in Appendix~\ref{sec:appa}.  

We calculate the contributions to Eq.~\ref{eq:zdecayamp} to $\mathcal{O}(\frac{1}{\Lambda^2})$, neglecting higher order  terms whose impact would be expected to be comparable to that of dimension-$8$ operators.  
The 1PI  loop amplitude $\mathcal{M}_1$ is given  in Appendix~\ref{sec:appb}. We use FeynArts~\cite{Hahn:2000kx} and FeynCalc~\cite{Mertig:1990an, Shtabovenko:2016sxi} to calculate loop amplitudes with the SMEFT package for FeynRules~\cite{Alloul:2013bka, Christensen:2009jx}. Explicit analytic expressions for the loop integrals have been computed using the FeynHelpers interface~\cite{Shtabovenko:2016whf} between FeynCalc and Package-X~\cite{Patel:2015tea}. As a check of our calculation, we demonstrate that the UV poles in Eq.~\ref{eq:zdecayamp} cancel completely in Appendix A. 

There are  IR divergences  arising from loops with massless photons, appearing in the fermion wave function renormalization and $\mathcal{M}_1$. We regulate these divergences with a photon mass,  $M_Z \beta$. We find,
\begin{align*}
   Re\biggl(
   Z_f\mathcal{M}_0+\mathcal{M}_1\biggr)&= \sqrt{G_\mu}
   {M_Z M_W^2\over {\sqrt[4]{2} \pi^2}}
   Q_f^2\log \beta
   (3+\log \beta)\sqrt{1-{M_W^2\over M_Z^2}}
   \biggl\{
   -G_{\mu} \sqrt{1-{M_W^2\over M_Z^2}} \left[ T_3^f-Q_f\biggl(1-{M_W^2\over M_Z^2} \biggr) \right] \\
&+\frac{C_{HWB}}{\sqrt{2} \Lambda^2} {M_W\over M_Z}\biggl[
2T_3^f-3Q_f\biggl(1-{M_W^2\over M_Z^2}\biggr)\biggr]\biggr\}\, . \numberthis
\label{eq:virtg}
\end{align*}

The above divergences give $\beta$-dependent terms in the decay width, which are in turn canceled by real photon emission
that contributes to both soft and collinear singularities. The SMEFT calculation of $Z \to f \bar{f} \gamma$ proceeds analogously to the well-known SM result~\cite{Ellis:1991qj} with additional terms proportional to $C_{HWB}$. The result is
\begin{eqnarray}
\left| \mathcal{M}(Z \to f \bar{f} \gamma) \right|^2 &=& {4G_\mu^2M_W^2M_Z^2 \over \pi^2}Q_f^2
\biggl\{
\biggl(T_3^f-Q_f s_W^2\biggr)\biggl[
   -\log (\beta ) (3+ \log \beta )
   +\frac{3
   E_0^2}{2 M_Z^2} \nonumber \\&&
   - \log \left(\theta_0\right)\biggl(
   \frac{3 E_0}{M_Z}+\log \left(\frac{M_Z}{2
   E_0}-1\biggr)-\frac{3}{4}\right)-\frac{E_0}{M_Z}+
   \frac{5 \pi ^2}{12}-\frac{87}{16}\biggr]\cdot 
    \biggl[-s_W^2\biggl (T_3^f-Q_f s_W^2\biggr)
    \nonumber\\ &&+{2c_Ws_W v ^2\over\Lambda^2}C_{HWB}(T_3^f-2Q_f s_W^2)\biggr]\biggr\}\, ,
\label{eq:photreal}
\end{eqnarray}
where $s_W^2=1-{M_W^2\over M_Z^2}, c_W={M_W\over M_Z}$ in Eq. \ref{eq:photreal} and
$\theta_0$ and $E_0$ are the angular and energy cutoff for observing the photon,  and depend on the detector sensitivities\cite{Kniehl:1991xe,Schwartz:2013pla}.  

After summing  Eq. \ref{eq:virtg} and the contributions from virtual and real photon emission, taking into account the fermion wave function renormalization, there is no $\beta$ dependence,
 verifying the cancellation of the IR divergences.
 In our numerical results below, we take $\theta_0 = 1^{\degree}$ and $E_0 = 1~\mathrm{GeV}$.

\subsection{Effective $Z$ Vertices}

From Eq.~\ref{eq:zdecayamp}, we obtain the contribution to the $Z \to f \bar{f}$ decay width from $C_{HWB}, C_{HW}, C_{HB}$ and $C_W$, still working to $\mathcal{O}(\frac{1}{\Lambda^2})$.
We write our result in terms of effective fermion couplings as
\begin{equation}
\Gamma(Z \to f_i \bar{f}_i) = \frac{G_{\mu} M_Z^3}{6 \sqrt{2} \pi} N_c (g_i^f)^2
\end{equation}
where $i = L,R$ indicates the fermion helicity and we neglect fermion masses. For a fermion with charge $Q_f$ and weak isospin $T_3^f$, the effective coupling is
\small
\begin{eqnarray}
g^f & =& (g^f)_{\mathrm{SM}} \bigg\{1 
+ \biggl({1~\mathrm{TeV}\over \Lambda}\biggr)^2{1\over D(Q_f,T_3^f) } \biggl[ -0.23 C_{HWB} Q_f^4+1.5 T_3^f C_{HWB} Q_f^3 \nonumber \\ &&+\left(-1.9
   C_{HWB} (T_3^{f})^2+0.15 C_{HB}+0.15 C_{HW}+11.0
   C_{HWB}+0.19 C_{W}\right) Q_f^2 \\ &&+ T_3^f (-0.67
   C_{HB}-0.69 C_{HW}-49.0 C_{HWB}-0.85 C_{W})
   Q_f+(T_3^{f})^2 (0.0084 C_{HB}+0.029 C_{HW}-0.23
   C_{HWB}+0.032 C_{W}) \nonumber
   \biggr]\biggr\}\, ,
\end{eqnarray}
\normalsize
where
\begin{equation}
D(Q_f,T_3^f)={Q_f^4 - 8.7 T_3^f Q_f^3 + \left( 17 (T_3^{f})^2 - 76 \right) Q_f^2 + 660 T_3^f Q_f - 1400 (T_3^{f})^2 }\, .
   \end{equation}
   The relatively large size of the $C_{HWB}$ coefficients is due to the fact that they contribute at tree level. 
For our numerical results we use, 
\begin{eqnarray}
G_\mu&=&1.1663787(6)\times 10^{-5}~\mathrm{GeV}^{-2}\nonumber \\
M_Z&=&91.1876\pm .0021~\mathrm{GeV}\nonumber \\
M_W&=&80.385\pm .015~\mathrm{GeV}\nonumber\\
M_H&=&125.09\pm 0.21\pm 0.11 ~\mathrm{GeV}\nonumber\\
M_t&=&173.1\pm0.6~\mathrm{GeV} \, .
\label{eq:input}
\end{eqnarray}

In particular, the SM fermion vertex couplings are
\begin{equation}
\begin{split}
g_L^\nu &= (g_L^\nu)_{\mathrm{SM}} \biggl[ 1 + \delta g_L^{Z\nu}
+ \biggl({1~\mathrm{TeV}\over \Lambda}\biggr)^2
\biggl\{ -6.0 \cdot 10^{-6} C_{HB}-2.1 \cdot 10^{-5} C_{HW} + 1.6 \cdot 10^{-4} C_{HWB} -2.3 \cdot 10^{-5} C_{W} \biggr\}\biggr] \\
g_L^e &= (g_L^e)_{\mathrm{SM}} \biggl[ 1 + \delta g_L^{Ze} 
+\biggl({1~\mathrm{TeV}\over \Lambda}\biggr)^2
\biggl\{ 0.0019 C_{HB}+0.0019 C_{HW} + 0.043 C_{HWB} +0.0023 C_{W} \biggr\}\biggr] \\
g_R^e &= (g_R^e)_{\mathrm{SM}} \biggl[ 1 + \delta g_R^{Ze} 
+\biggl({1~\mathrm{TeV}\over \Lambda}\biggr)^2
\biggl\{-0.0020 C_{HB}-0.0020 C_{HW} - 0.033 C_{HWB} -0.0025 C_{W} \biggr\}\biggr] \\
g_L^u &= (g_L^u)_{\mathrm{SM}} \biggl\{ 1 + \delta g_L^{Zu} + 
\biggl({1~\mathrm{TeV}\over \Lambda}\biggr)^2
\biggl\{9.3 \cdot 10^{-4} C_{HB}+ 9.3 \cdot 10^{-4} C_{HW} + 0.021 C_{HWB} +0.0011 C_{W} \biggr\}\biggr] \\
g_R^u &= (g_R^u)_{\mathrm{SM}} \biggl[ 1 + \delta g_R^{Zu}
+\biggl({1~\mathrm{TeV}\over \Lambda}\biggr)^2
\biggl\{ -0.0020 C_{HB}-0.0020 C_{HW} - 0.034 C_{HWB} -0.0025 C_{W} \biggr\}\biggr] \\
g_L^d &= (g_L^d)_{\mathrm{SM}} \biggl[1 + \delta g_L^{Zd} + 
\biggl({1~\mathrm{TeV}\over \Lambda}\biggr)^2
\biggl\{3.7 \cdot 10^{-4} C_{HB}+ 3.6 \cdot 10^{-4} C_{HW} + 0.0080 C_{HWB} + 4.5 \cdot 10^{-4} C_{W} \biggr\}\biggr] \\
g_R^d &= (g_R^d)_{\mathrm{SM}} \biggl[ 1 + \delta g_R^{Zd}+
\biggl({1~\mathrm{TeV}\over \Lambda}\biggr)^2
\biggl\{ -0.0020 C_{HB}-0.0020 C_{HW} - 0.034 C_{HWB} -0.0025 C_{W} \biggr\}\biggr]\, .
\label{eq:couplings}
\end{split}
\end{equation}
For $b_L$, the coefficient in front of $C_W$ is $2.7 \cdot 10^{-4}$ rather than $4.5 \cdot 10^{-4}$ because of top mass effects.  The tree level contributions
of $C_{HWB}$ are contained in the $\delta g_{L,R}^{Zf}$ contributions as given in Eq. \ref{eq: mapcoef}.

\begin{figure}
\centering
\includegraphics[width=0.49\textwidth]{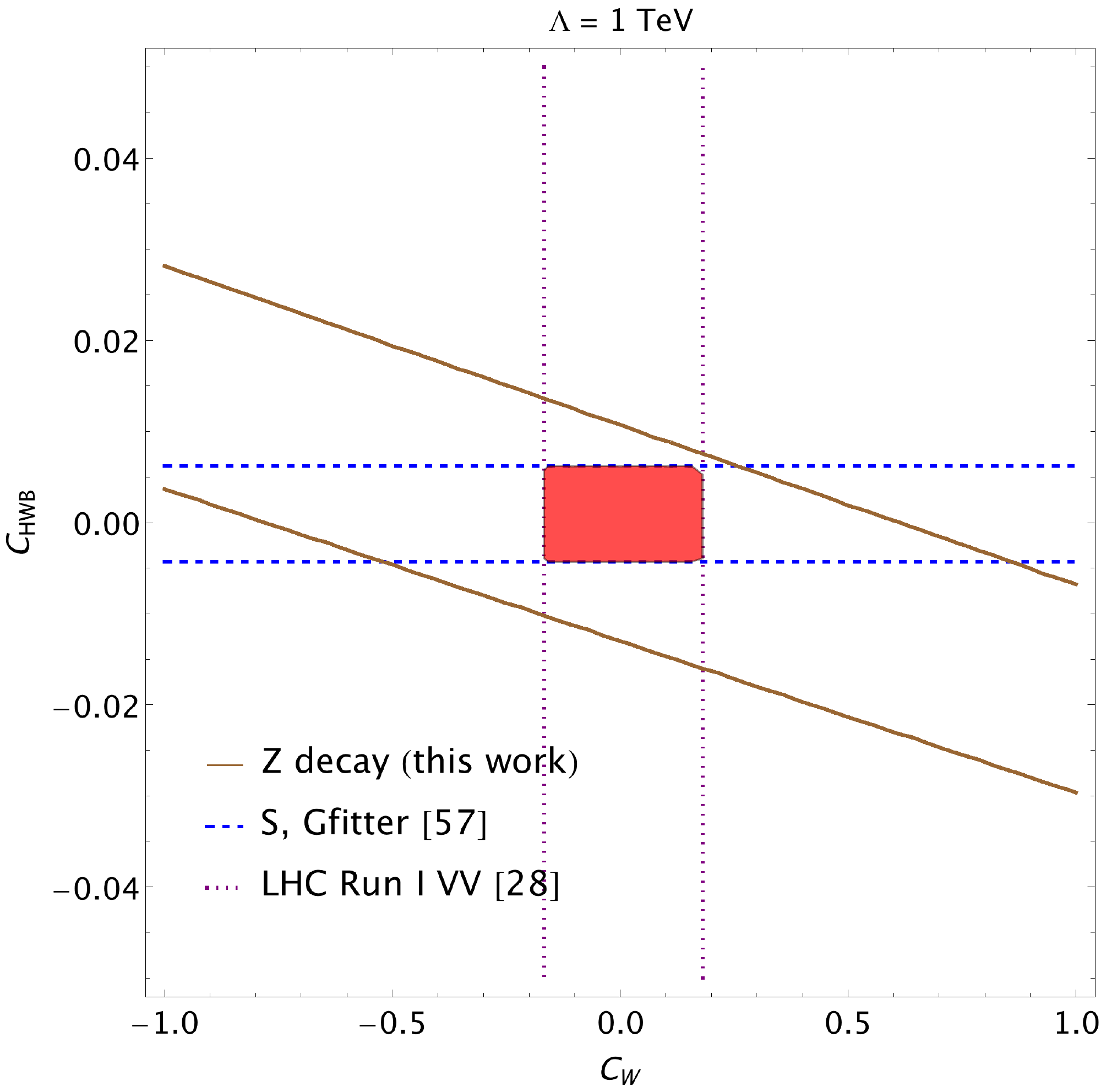}
\caption{Limits on the coefficients of the operators $\mathcal{O}_W$ and $\mathcal{O}_{HWB}$ which contribute to $Z$ decay at one loop and tree level, respectively. All other operators are set to zero, and the region between the solid brown lines is allowed by $Z$ pole measurements given our calculation. For comparison, the region between the dashed blue lines is allowed by the same LEP precision data considering only the impact of the operator $\mathcal{O}_{HWB}$ that modifies the $S$ parameter. The region between the magenta dotted lines is allowed by measurements of $VV$ production at the LHC, to which $\mathcal{O}_W$ contributes at tree level. The region in the plane that is allowed by all measurements is shown in red.}
\label{fig:opplots}
\end{figure}

\begin{figure}
\centering
\includegraphics[width=0.49\textwidth]{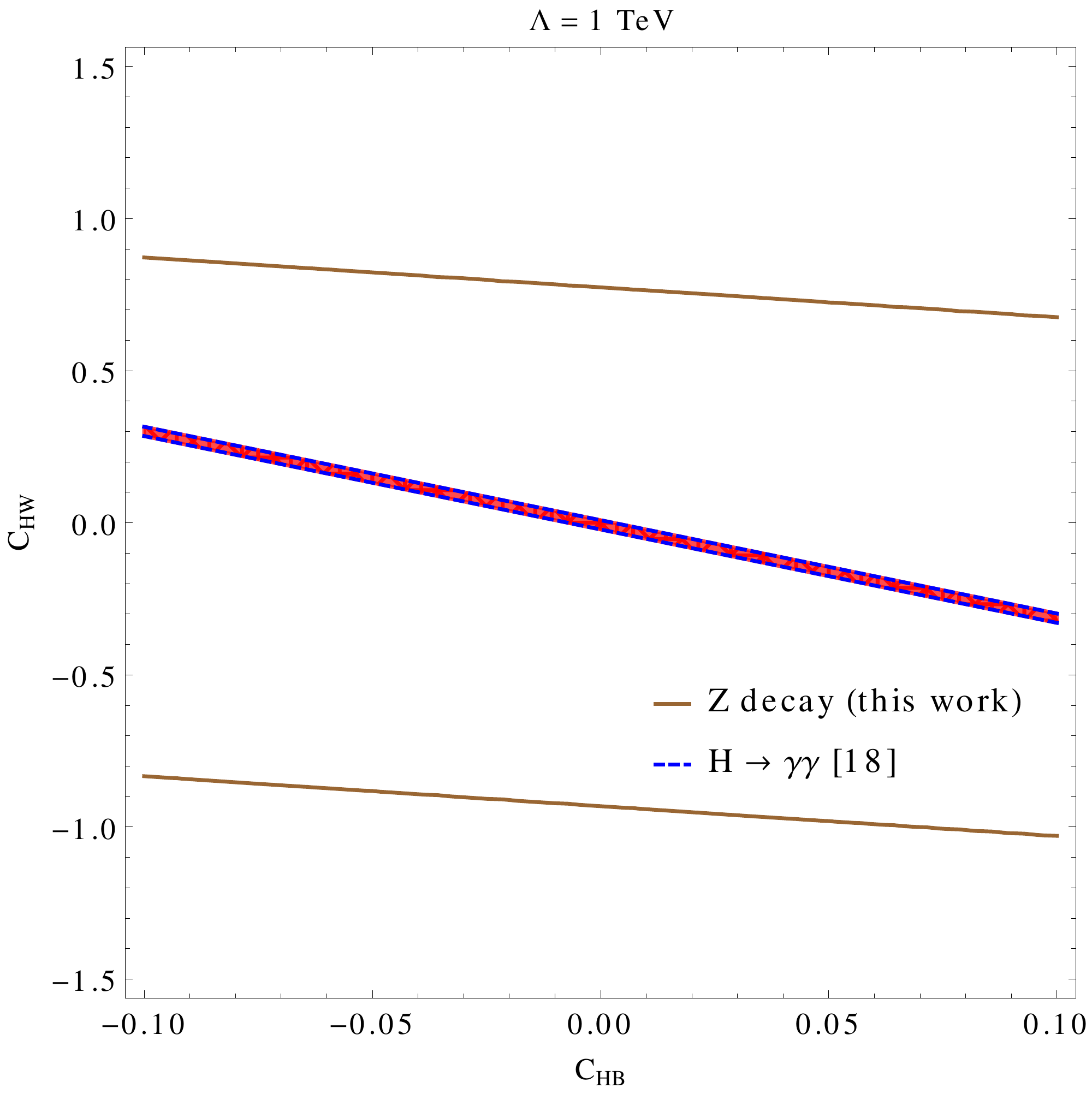}
\caption{Limits on the coefficients of the operators $\mathcal{O}_{HB}$ and $\mathcal{O}_{HW}$ which contribute to $Z$ decay at one loop. The solid brown lines are as in Fig.~\ref{fig:opplots}. (The allowed 
region is between the lines).  For comparison, the region allowed by the more constraining measurement of the $H \to \gamma \gamma$ signal strength is shown in red between the blue dashed lines.} 
\label{fig:ggpplots}
\end{figure}

\begin{table}
\centering
\newcolumntype{C}{>{\centering\arraybackslash}X}
\begin{tabularx}{\textwidth}{|c|c|c||C|C|C|C|C|C|C|}
\hline
\multirow{2}{*}{Parameter} & \multirow{2}{*}{SM prediction} & \multirow{2}{*}{Measurement} & \multicolumn{7}{c|}{Correlations} \\
\cline{4-10}
&&& $g_L^\nu$ & $g_L^\ell$ & $g_L^b$ & $g_L^c$ & $g_R^\ell$ & $g_R^b$ & $g_R^c$ \\
\hline
\hline
$g_L^\nu$ & $0.50199 \pm 0.00020$ & $0.50075 \pm 0.00077$ & 1.00 &&&&&& \\
$g_L^\ell$ & $-0.26919 \pm 0.00020$ & $-0.26939 \pm 0.00022$ & -0.32 & 1.00 &&&&& \\
$g_L^b$ & $-0.42114 \pm 0.00045$ & $-0.4182 \pm 0.0015$ & 0.05 & -0.27 & 1.00 &&&& \\
$g_L^c$ & $0.34674 \pm 0.00017$ & $0.3453 \pm 0.0036$ & -0.02 & 0.04 & -0.09 & 1.00 &&& \\
$g_R^\ell$ & $0.23208 \pm 0.00018$ & $0.23186 \pm 0.00023$ & 0.25 & 0.34 & -0.37 & 0.07 & 1.00 && \\
$g_R^b$ & $0.077420 \pm 0.000052$ & $0.0962 \pm 0.0063$ & 0.00 & -0.33 & 0.88 & -0.14 & -0.35 & 1.00 & \\
$g_R^c$ & $-0.15470 \pm 0.00011$ & $-0.1580 \pm 0.0051$ & 0.00 & 0.08 & -0.17 & 0.30 & 0.08 & -0.13 & 1.00 \\
\hline
\end{tabularx}
\caption{LEP and SLD measurements of the effective $Z$ couplings to fermions.}
\label{tab:lep}
\end{table}

These effective couplings are bounded by LEP measurements at the $Z$ pole. We proceed to take the limits of~\cite{ALEPH:2005ab} on the $Z$-fermion couplings to constrain the SMEFT operators. We minimize a $\chi^2$ function constructed using the LEP measurements of the quantities $(g_L^\nu, g_L^e, g_R^e, g_L^u, g_R^u, g_L^b, g_R^b)$ and their correlations
 shown in Table \ref{tab:lep},
 including the uncertainties on the input parameters. While $Z$ pole measurements constrain all of the operators in Eq.~\ref{eq:couplings}, we focus on the implications of our calculation for the operators $\mathcal{O}_W$, $\mathcal{O}_{HB}$ and $\mathcal{O}_{HW}$, which do not contribute to $Z$ decay at tree level. 
 We seek to minimize the quantity
\begin{equation}
(\chi^2)_{\mathrm{LEP}} = \left(\vec{g}_{\mathrm{SMEFT}} - \vec{g}_{\mathrm{exp}}\right)^\mathrm{T} V^{-1} \left(\vec{g}_{\mathrm{SMEFT}} - \vec{g}_{\mathrm{exp}} \right)\, ,
\end{equation}
where $\vec{g} = \left( g_L^\nu, g_L^\ell, g_L^b, g_L^c, g_R^\ell, g_R^b, g_R^c \right)$ and $V$ is the covariance matrix constructed from the errors and correlations above. We use Eq.~\ref{eq:couplings} together with the SM predictions of Table~\ref{tab:lep} to calculate $\vec{g}_{\mathrm{SMEFT}}$. Since we set light fermion masses to zero in our SMEFT analysis, the effective couplings for the down (up) quark apply equally to the $b$ ($c$) quark, with the exception of the $b_L$ for which top quark corrections apply as specified below Eq.~\ref{eq:couplings}.

 In Fig.~\ref{fig:opplots}, we show the resulting 90\% CL limits in 2-dimensional planes of the coefficients of these operators along with that of $\mathcal{O}_{HWB}$, which affects electroweak couplings at tree level. The coefficients of all other operators are set to zero. 
 We compare our results to processes in which the SMEFT operators contribute at tree level. The limit of ~\cite{Alves:2018nof}, set using LHC Run I data~\cite{Khachatryan:2015sga,Aad:2016wpd,Aad:2016ett,Khachatryan:2016poo}, is converted in our notation to
\begin{equation}
-0.17 < C_W \left( \frac{1~\mathrm{TeV}}{\Lambda} \right)^2 < 0.18\, .
\end{equation}
 For $\mathcal{O}_W$, we use the limits of~\cite{Alves:2018nof} obtained by  using 8 TeV LHC gauge boson pair production in leptonic final states~\cite{Khachatryan:2015sga,Aad:2016wpd,Aad:2016ett,Khachatryan:2016poo}. 

For $\mathcal{O}_{HB}$ and $\mathcal{O}_{HW}$, we use limits~\cite{Dawson:2018liq} from the calculation of $H \to \gamma \gamma$~\cite{Hartmann:2015aia,Dedes:2018seb,Dawson:2018liq} in the SMEFT, as compared to measurements of $H \to \gamma \gamma$ at Run 1 and 2 of the LHC~\cite{Khachatryan:2016vau,Aaboud:2018xdt,Sirunyan:2018ouh}. 
The SMEFT calculation of $H \to \gamma \gamma$~\cite{Dawson:2018liq} gives,
\begin{eqnarray}
\mu_{\gamma\gamma}&=&{\Gamma(H\rightarrow \gamma\gamma)\over \Gamma(H\rightarrow \gamma\gamma)\mid_{SM}}\nonumber \\
&=&1-40.15{C_{HB}} \left( \frac{1~\mathrm{TeV}}{\Lambda} \right)^2-13.08{C_{HW}} \left( \frac{1~\mathrm{TeV}}{\Lambda} \right)^2
+22.40{C_{HWB}} \left( \frac{1~\mathrm{TeV}}{\Lambda} \right)^2\, .
\end{eqnarray}
Then, using the average $\mu_{\gamma\gamma} = 1.09 \pm 0.10$ of current LHC Higgs measurements~\cite{Khachatryan:2016vau,Aaboud:2018xdt,Sirunyan:2018ouh}, we find,
\begin{eqnarray}
-0.003 < \mid {C_{HB}} \left( \frac{1~\mathrm{TeV}}{\Lambda} \right)^2+.33{C_{HW}} \left( \frac{1~\mathrm{TeV}}{\Lambda} \right)^2\
-.55{C_{HWB}} \left( \frac{1~\mathrm{TeV}}{\Lambda} \right)^2\mid < 0.007\, ,
\end{eqnarray}
or taking only one non-zero coupling at a time with the conservative bound $|\mu_{\gamma\gamma} - 1| < 0.29$, 
\begin{equation}
\begin{split}
\left|C_{HB} \left( \frac{1~\mathrm{TeV}}{\Lambda} \right)^2\right| &< 0.007 \\
\left|C_{HW} \left( \frac{1~\mathrm{TeV}}{\Lambda} \right)^2\right| &< 0.02\, . 
\end{split}
\end{equation}
$\mathcal{O}_{HWB}$ corresponds to the oblique parameter $S$~\cite{Peskin:1990zt,Peskin:1991sw}, whose limit we take from the Gfitter collaboration\cite{Haller:2018nnx}
of $S = 0.04 \pm 0.11$ to set the $2 \sigma$ bound,
\begin{equation}
-0.004 < C_{HWB} < 0.006 \, .
\end{equation}

The existing bounds in Fig.~\ref{fig:ggpplots} are stronger than those that we obtain directly from $Z$ pole measurements. Nevertheless, they provide complementary information, and in particular in  Fig. \ref{fig:ggpplots}, the interplay between the limits on $C_W$ and $C_{HWB}$ demonstrates the power of electroweak precision measurements to constrain couplings that only contribute at loop level. In the case of the operators $C_{HB}$ and $C_{HW}$ which directly affect $H \to \gamma \gamma$, Higgs precision is already significantly more effective than $Z$ pole measurements in setting limits, due to the loop suppression of these operators' contributions to $Z$ decay.

\section{Conclusions}
\label{sec:concl}

Precision measurements of electroweak physics will eventually necessitate higher order calculations of BSM contributions. The SMEFT framework takes a general approach to potential new UV physics by parametrizing its effects in terms of higher dimension operators involving the SM fields. In this work, we have furthered the applicability of the SMEFT to probe new physics by considering the one loop corrections to $Z$ decay from operators which contribute to gauge boson production.

While the contributions of the operators $\mathcal{O}_W$, $\mathcal{O}_{HB}$ and $\mathcal{O}_{HW}$ are small relative to those of the operators that modify the $Z$ coupling to fermions at tree level, the relative size of all of the SMEFT operators is fixed by the new physics. In particular, integrating out a heavy SM singlet scalar could naturally give these operators without changing the leading $Z$ couplings to the fermions~\cite{deBlas:2017xtg}. In such a scenario, it would be essential to have the higher order contributions of the BSM physics to all possible processes. In this regard our calculation provides a useful prediction, relating the effects of new physics in $Z$ decay to those in other electroweak processes provided the states responsible for deviations from the SM are heavy enough to be integrated out.

A full calculation of $Z$ decay at one loop in the SMEFT would provide even more complete information about the influence of higher dimensional operators on $Z$ physics. With this as well as other higher order calculations of electroweak processes, in the future a global fit at NLO  in the SMEFT could be performed to bound the sizes of all possible dimension-$6$ SMEFT operators.
\section*{Acknowledgements}
We thank Ayres Freitas and Pier Paolo Giardino for useful discussions. SD is supported by the U.S.~Department of Energy under Grant Contract DE-SC0012704. AI is supported by the U.S.~Department of Energy under Grant Contract DE-SC0015634 and by PITT PACC.
\appendix
\section{UV poles}
\label{sec:poles}

The cancellation of UV poles follows from the individual contributions:
Numerically with $\Lambda = 1~\mathrm{TeV}$, the pieces are as follows.
\begin{align*}
 \frac{\partial M_0}{\partial C_{HWB}} \delta C_{HWB}
   &= \frac{1}{\epsilon} \Big\{ Q_f \left(\left(2.7\times 10^{-5}\right) C_{HB}+\left(2.7\times
   10^{-5}\right) C_{HW}+\left(4.6\times 10^{-4}\right)
   C_{HWB}+\left(2.6\times 10^{-5}\right) C_{W}\right) \Big\} + \mathcal{O}(\epsilon^0) \\
 \frac{\partial M_0}{\partial G_{\mu}} \delta G_{\mu} &= \frac{1}{\epsilon} \Big\{ \left(2.1\times
   10^{-5}\right) \left(Q_f (C_{HWB}+8.8)-40
   T_3^f\right) (\xi -5.5) \Big\} + \mathcal{O}(\epsilon^0) \\
 \frac{\partial M_0}{\partial M_Z^2} \delta M_Z^2 &= \frac{1}{\epsilon} \Big\{ Q_f
   \big(\left(-1.4\times 10^{-4}\right) \xi  C_{HW}-\left(3.2\times
   10^{-4}\right) C_{HW}-\left(6.7\times 10^{-4}\right)
   C_{HWB}+\left(7.7\times 10^{-4}\right) C_{W} \\&\quad +C_{HB}
   \left(\left(-3.9\times 10^{-5}\right) \xi -9.1\times
   10^{-5}\right)-\left(1.4\times 10^{-6}\right) C_{HWB} \xi
   -\left(1.5\times 10^{-3}\right) \xi +1.2\times 10^{-2}\big) \\&\quad +T_3^f
   \big(\left(7.7\times 10^{-5}\right) \xi  C_{HW}+\left(1.8\times
   10^{-4}\right) C_{HW}+\left(5.3\times 10^{-5}\right)
   C_{HWB}-\left(4.3\times 10^{-4}\right) C_{W} \\&\quad +C_{HB}
   \left(\left(2.2\times 10^{-5}\right) \xi +5.1\times
   10^{-5}\right)+\left(4.1\times 10^{-5}\right) C_{HWB} \xi
   +\left(8.2\times 10^{-4}\right) \xi -6.5\times 10^{-3}\big) \Big\} + \mathcal{O}(\epsilon^0) \\
 \frac{\partial M_0}{\partial M_W^2} \delta M_W^2 &= \frac{Q_f}{\epsilon} \Big\{
  \left(-9.2\times 10^{-4}\right) C_{W}+C_{HWB}
   \left(5.3\times 10^{-4}-\left(5.2\times 10^{-5}\right) \xi
   \right)+C_{HW} \left(\left(1.5\times 10^{-4}\right) \xi +3.0\times
   10^{-4}\right) \\&\quad +\left(1.3\times 10^{-3}\right) \xi -1.2\times
   10^{-2} \Big\} + \mathcal{O}(\epsilon^0) \\
 -M_{\gamma }\frac{\Pi_{\gamma Z}\left(M_Z^2\right)}{M_Z^2} &= \frac{Q_f}{\epsilon} \Big\{
   \left(3.7\times 10^{-5}\right) \xi  C_{HWB}-\left(2.3\times
   10^{-4}\right) C_{HWB}+\left(1.9\times 10^{-4}\right)
   C_{W}+C_{HW} \big(\left(-3.5\times 10^{-5}\right) \xi
   \\&\quad -5.2\times 10^{-5}\big)+C_{HB} \left(\left(3.5\times
   10^{-5}\right) \xi +5.2\times 10^{-5}\right)-\left(1.7\times
   10^{-4}\right) \xi +1.3\times 10^{-3} \Big\} + \mathcal{O}(\epsilon^0) \\
 \frac{Z_Z}{2}M_0 &= \frac{1}{\epsilon} \Big\{ T_3^f \big(\left(-7.7\times 10^{-5}\right) \xi 
   C_{HW}-\left(1.8\times 10^{-4}\right) C_{HW}-\left(1.3\times
   10^{-4}\right) C_{HWB}+\left(3.2\times 10^{-4}\right)
   C_{W} \\&\quad +C_{HB} \left(\left(-2.2\times 10^{-5}\right) \xi
   -5.1\times 10^{-5}\right)-\left(6.7\times 10^{-5}\right) C_{HWB}
   \xi -\left(7.8\times 10^{-4}\right) \xi -3.1\times 10^{-4}\big) \\&\quad +Q_f
   \big(\left(1.7\times 10^{-5}\right) \xi  C_{HW}+\left(4.0\times
   10^{-5}\right) C_{HW}+\left(2.1\times 10^{-5}\right)
   C_{HWB}-\left(7.0\times 10^{-5}\right) C_{W} \\&\quad +C_{HB}
   \left(\left(4.9\times 10^{-6}\right) \xi +1.1\times
   10^{-5}\right)-\left(4.7\times 10^{-6}\right) C_{HWB} \xi
   +\left(1.7\times 10^{-4}\right) \xi +6.9\times 10^{-5}\big) \Big\} + \mathcal{O}(\epsilon^0) \\
 Z_f M_0 &= \frac{\xi}{\epsilon} \Big\{ Q_f^3 \left(1.3\times 10^{-4}-\left(4.3\times 10^{-5}\right)
   C_{HWB}\right) +Q_f^2 T_3^f \left(\left(1.9\times 10^{-4}\right)
   C_{HWB}-8.3\times 10^{-4}\right) \\&\quad +Q_f T_3^{f2} \left(2.6\times
   10^{-3}-\left(3.0\times 10^{-4}\right) C_{HWB}\right)
   -\left(1.6\times 10^{-3}\right) T_3^f \Big\} + \mathcal{O}(\epsilon^0) \\
 M_1 &= \frac{1}{\epsilon} \Big\{ Q_f^3 \left(\left(4.3\times 10^{-5}\right) C_{HWB}-1.3\times
   10^{-4}\right) \xi +Q_f T_3^{f2} \big(\left(3.0\times 10^{-4}\right)
   C_{HWB}-2.6\times 10^{-3}\big) \xi \\&\quad +T_3^f
   \big(\left(8.3\times 10^{-4}\right) \xi  Q_f^2+\left(1.2\times
   10^{-4}\right) C_{W}+\left(2.4\times 10^{-3}\right) \xi
   +C_{HWB} \big(\left(-1.9\times 10^{-4}\right) \xi 
   Q_f^2 \\&\quad +\left(2.5\times 10^{-5}\right) \xi +7.6\times
   10^{-5}\big)+2.3\times 10^{-3}\big) \Big\} + \mathcal{O}(\epsilon^0)
\end{align*}

The sum vanishes for any given fermion.

\section{$2$-point functions}
\label{sec:appa}
In this appendix, we show the two-point functions in $R_\xi$ gauge  due to the SMEFT operators that also contribute to gauge boson pair production. Previous results for the gauge boson two-point functions in other operator bases appear in~\cite{Alam:1997nk, Chen:2013kfa}.

In $D=4-2\epsilon$ dimensions, the two-point function for a massless fermion with weak isospin $T_3^f$ and charge $Q_f$ is
\begin{align*}
\Sigma(p) &= \frac{1}{8 \pi ^2} \Bigg( \frac{2 \sqrt{2} A_0(M_W^2) G_{\mu} T_3^{f2} \left((D-2) M_W^2-p^2\right)}{p^2}+2 \sqrt{2} A_0(M_W^2 \xi) G_{\mu} T_3^{f2} \\&\quad +\frac{1}{M_Z^4 p^2} \bigg(A_0(M_Z^2)
   \left((D-2) M_Z^2-p^2\right) \left(M_W^2 Q_f+M_Z^2 (T_3^f-Q_f)\right) \Big(\sqrt{2} G_{\mu} \left(M_W^2 Q_f+M_Z^2
   (T_3^f-Q_f)\right) \\&\quad +2  M_W Q_f C_{HWB} \sqrt{M_Z^2-M_W^2}\Big)\bigg) \\&\quad -\frac{A_0(M_Z^2 \beta) (D-2) M_W^2 Q_f^2
   \left(\sqrt{2} G_{\mu} (M_W-M_Z) (M_W+M_Z)+2  M_W C_{HWB} \sqrt{M_Z^2-M_W^2}\right)}{M_Z^2
   p^2} \\&\quad +\frac{A_0(M_Z^2 \xi) \left(M_W^2 Q_f+M_Z^2 (T_3^f-Q_f)\right) \left(\sqrt{2} G_{\mu} \left(M_W^2 Q_f+M_Z^2 (T_3^f-Q_f)\right)+2
    M_W Q_f C_{HWB} \sqrt{M_Z^2-M_W^2}\right)}{M_Z^4} \\&\quad -\frac{B_0(0, M_Z^2 \beta, M_Z^2 \beta) M_W^2 (\xi -1) Q_f^2 \left(\sqrt{2}
   G_{\mu} (M_W-M_Z) (M_W+M_Z)+2  M_W C_{HWB} \sqrt{M_Z^2-M_W^2}\right)}{M_Z^2} \\&\quad +\frac{2 \sqrt{2}
   B_0(p^2, 0, M_W^2) G_{\mu} T_3^{f2} \left(p^2-M_W^2\right) \left((D-2) M_W^2+p^2\right)}{p^2}-2 \sqrt{2} B_0(p^2, 0, M_W^2 \xi) G_{\mu} T_3^{f2}
   \left(p^2-M_W^2 \xi \right) \\&\quad -\frac{1}{M_Z^4 p^2}\bigg(B_0(p^2, 0, M_Z^2) (M_Z-p) (M_Z+p) \left((D-2) M_Z^2+p^2\right) \left(M_W^2 Q_f+M_Z^2
   (T_3^f-Q_f)\right) \Big(\sqrt{2} G_{\mu} \big(M_W^2 Q_f \\&\quad +M_Z^2 (T_3^f-Q_f)\big)+2  M_W Q_f C_{HWB}
   \sqrt{M_Z^2-M_W^2}\Big)\bigg) \\&\quad +\frac{1}{M_Z^2
   p^2}\bigg(B_0(p^2, 0, M_Z^2 \beta) M_W^2 Q_f^2 \left(\beta  (D-2) M_Z^2-p^2 (D+\xi -3)\right)
   \Big(\sqrt{2} G_{\mu} (M_W-M_Z) (M_W+M_Z) \\&\quad +2  M_W C_{HWB} \sqrt{M_Z^2-M_W^2}\Big)\bigg) \\&\quad +\frac{1}{M_Z^4} \bigg(B_0(p^2, 0, M_Z^2 \xi) \left(M_Z^2 \xi -p^2\right) \left(M_W^2 Q_f+M_Z^2 (T_3^f-Q_f)\right) \Big(\sqrt{2} G_{\mu} \left(M_W^2
   Q_f+M_Z^2 (T_3^f-Q_f)\right) \\&\quad +2  M_W Q_f C_{HWB} \sqrt{M_Z^2-M_W^2}\Big)\bigg) +\frac{1}{M_Z^2} \bigg(C_0(0, p^2, p^2, M_Z^2 \beta, M_Z^2 \beta, 0)
   M_W^2 (\xi -1) Q_f^2 \left(p^2-\beta  M_Z^2\right) \\&\quad \times \left(\sqrt{2} G_{\mu} (M_W-M_Z) (M_W+M_Z)+2  M_W
   C_{HWB} \sqrt{M_Z^2-M_W^2}\right)\bigg) \Bigg) \numberthis
\end{align*}
We have regulated IR divergences with a photon of mass $M_Z \beta$, and use standard FeynCalc notation~\cite{Mertig:1990an} for the Passarino-Veltman functions.

This leads to the wave function renormalization
\begin{align*}
Z_f &= \frac{1}{8 \pi ^2 \epsilon } \Bigg(\frac{1}{M_Z^2} \bigg(Q_f^2 \left(\sqrt{2} G_{\mu} (M_W-M_Z) (M_W+M_Z)+2  M_W C_{HWB} \sqrt{M_Z^2-M_W^2}\right) \big(2
   \xi  \epsilon  \big(M_W^2 (\log (\beta )-\log (\xi )) \\&\quad +M_Z^2 (\log (M_Z^2 / \mu^2)+\log (\pi  \xi ))\big)+2 M_W^2 (\xi -1)
   \epsilon +M_Z^2 ((2 (\gamma -1) \xi +3) \epsilon -2 \xi )\big)\bigg) \\&\quad -2 Q_f T_3^f \left(\sqrt{2} G_{\mu} (M_W-M_Z)
   (M_W+M_Z)+ M_W C_{HWB} \sqrt{M_Z^2-M_W^2}\right) (2 \xi  \epsilon  (2 \log (M_Z^2 / \mu^2)+\log (\pi  \xi ))-2 \xi \\&\quad +(2 (\gamma -1) \xi +3) \epsilon )-\sqrt{2} G_{\mu} T_3^{f2} \big(\left(2 M_W^2+M_Z^2\right) (2 \xi  \epsilon  (\log (\pi 
   \xi ))+2 \xi  ((\gamma -1) \epsilon -1)+3 \epsilon )+4 M_W^2 \xi  \epsilon  \log (M_W^2 / \mu^2) \\&\quad +2 M_Z^2 \xi  \epsilon  \log
   (M_Z^2 / \mu^2)\big)\Bigg) \numberthis
\end{align*}
For the $b_L$, there are corrections proportional to the top mass, leading to an additional wave function renormalization which in Feynman gauge is
\begin{equation}
Z_{b_L} = Z_f \left( Q_f = -\frac{1}{3}, T_3^f = -\frac{1}{2} \right) + \frac{G_{\mu} M_T^2 \left( 2 \epsilon \log \left(\frac{\mu^2}{M_W^2}\right) - 2 \gamma \epsilon + \epsilon - 2 \epsilon \log \pi + 2 \right)}{16 \sqrt{2} \pi^2 \epsilon}
\end{equation}

The transverse $W$ two-point function is
\begin{align*}
\Pi_{WW}^T(p^2) &= \frac{1}{16 p^2 \pi ^2} \Bigg(-\frac{18 \sqrt{2} B_0(p^2, 0, 0) (D-2) G_{\mu} M_W^2
   p^4}{D-1} -\frac{1}{(D-1) M_Z^2} \bigg(C_0(0, p^2, p^2, 0, 0, M_W^2 \xi) \left(M_W^4-p^4\right)
   \\&\quad \times \left(\sqrt{2} G_{\mu} (M_W-M_Z) (M_W+M_Z)+2
    M_W \sqrt{M_Z^2-M_W^2} C_{HWB}\right)
   (\xi -1) \left(p^2-M_W^2 \xi \right)^2 \bigg) \\&\quad -\frac{6
   \sqrt{2} A_0(M_T^2) G_{\mu} M_W^2 \left(M_T^2-(D-2)
   p^2\right)}{D-1}+\frac{6 \sqrt{2} B_0(p^2, 0, M_T^2) G_{\mu} M_W^2
   (M_T-p) (M_T+p) \left(M_T^2+(D-2)
   p^2\right)}{D-1} \\&\quad -\frac{1}{D-1} \bigg(B_0(p^2, M_H^2, M_W^2) M_W^2 \Big(8 (D-1) 
   \left(-M_H^2+M_W^2+p^2\right) C_{HW} p^2 \\&\quad +\sqrt{2}
   G_{\mu} \left(p^4-2 \left(M_H^2+(3-2 D) M_W^2\right)
   p^2+\left(M_H^2-M_W^2\right)^2\right)\Big) \bigg) +\frac{1}{D-1} \bigg(
   A_0(M_H^2) \Big(\sqrt{2} G_{\mu} M_W^2
   \big(M_H^2-M_W^2 \\&\quad -(D-2) p^2\big)-2 (D-1)  p^2
   \left(4 M_W^2+p^2\right)
   C_{HW}\Big)\bigg)-\frac{1}{(D-1) M_Z^4} \bigg(B_0(p^2, M_W^2, M_Z^2) M_W
   (M_W+M_Z-p) (M_W+ \\&\quad M_Z+p)
   \left((M_W-M_Z)^2-p^2\right) \Big(36 \sqrt[4]{2} (D-2)
   \sqrt{G_{\mu}}  M_W^2 M_Z^2 C_{W}
   p^2+\sqrt{2} G_{\mu} M_W \big(M_W^4 \\&\quad +2 (2 D-3)
   \left(M_Z^2+p^2\right) M_W^2+M_Z^4+p^4+2 (2 D-3)
   M_Z^2 p^2\big)+2  \sqrt{M_Z^2-M_W^2}
   \big(M_W^4+(2 D-3) \left(M_Z^2+2 p^2\right)
   M_W^2 \\&\quad +p^4+(2 D-3) M_Z^2 p^2\big)
   C_{HWB}\Big)\bigg)+\frac{1}{(D-1)
   M_Z^2} \bigg(C_0(0, p^2, p^2, 0, 0, M_W^2)
   \left(M_W^2-p^2\right)^2 \big(M_W^4 \\&\quad +2 (2 D-3) p^2
   M_W^2+p^4\big) \left(\sqrt{2} G_{\mu}
   \left(M_Z^2-M_W^2\right)-2  M_W
   \sqrt{M_Z^2-M_W^2} C_{HWB}\right) (\xi -1) \bigg) \\&\quad +\frac{1}{(D-1)
   M_Z^4} \bigg(B_0(p^2, M_W^2 \xi, M_Z^2 \xi) M_W
   \left(M_W^4-p^4\right) \left(\sqrt{2} G_{\mu} M_W+2
    \sqrt{M_Z^2-M_W^2} C_{HWB}\right) \\&\quad \times
   \left(p^4-2 \left(M_W^2+M_Z^2\right) \xi 
   p^2+\left(M_W^2-M_Z^2\right)^2 \xi ^2\right)\bigg)+\frac{1}{(D-1)
   M_Z^4} \bigg( B_0(p^2, M_W^2, M_Z^2 \xi) M_W
   \left(M_W^2-p^2\right)^2 \\&\quad \times \left(\sqrt{2} G_{\mu} M_W+2
    \sqrt{M_Z^2-M_W^2} C_{HWB}\right)
   \left(M_W^4+\left((4 D-6) p^2-2 M_Z^2 \xi \right)
   M_W^2+\left(p^2-M_Z^2 \xi \right)^2\right) \bigg) \\&\quad -\frac{1}{D-1} \bigg( A_0(M_W^2 \xi) \Big(2  \left(2 (D-1)
   C_{HW} p^4+M_W \sqrt{M_Z^2-M_W^2} C_{HWB}
   p^2-M_W^3 \sqrt{M_Z^2-M_W^2}
   C_{HWB}\right) \\&\quad +\sqrt{2} G_{\mu} \left((\xi -1) M_W^4-2 p^2
   (D+\xi -3) M_W^2+p^4 (4 D+\xi -7)\right)\Big) \bigg) -\frac{1}{(1-D) M_Z^2} \bigg(2
   B_0(p^2, 0, M_W^2) \\&\quad \times \Big(18 \sqrt[4]{2} (D-2) \sqrt{G_{\mu}} 
   M_W (M_W-M_Z) (M_W+M_Z) (M_W-p)^2 p^2
   C_{W} (M_W+p)^2 \\&\quad +\sqrt{2} G_{\mu} (M_W-M_Z)
   (M_W+M_Z) \left(M_W^2+p^2\right) \left((2 D+\xi -4)
   M_W^4-2 p^2 (4 D+\xi -6) M_W^2+p^4 (2 D+\xi
   -4)\right) \\&\quad + M_W \sqrt{M_Z^2-M_W^2}
   \left(M_W^2+p^2\right) C_{HWB} \left((2 D+2 \xi -5)
   M_W^4-2 p^2 (6 D+2 \xi -9) M_W^2+p^4 (2 D+2 \xi
   -5)\right)\Big)\bigg) \\&\quad -\frac{1}{(D-1)
   M_Z^2} \bigg(2 B_0(p^2, 0, M_W^2 \xi)
   (M_W-p) (M_W+p) \bigg(\sqrt{2} G_{\mu}
   (M_W-M_Z) (M_W+M_Z) \Big(\xi  M_W^4-p^2
   \big(\xi ^2+2 \xi \\&\quad +2 D-3\big) M_W^2-p^4 (2 D+\xi
   -4)\Big)- M_W \sqrt{M_Z^2-M_W^2}
   C_{HWB} \big((\xi -2) \xi  M_W^4+2 p^2 \left(\xi ^2+D (\xi
   +2)-3\right) M_W^2 \\&\quad +p^4 (2 D+2 \xi -5)\big)\bigg)\bigg)+\frac{1}{(D-1) M_Z^4} \bigg(A_0(M_Z^2 \xi) \Big(\sqrt{2} G_{\mu} \big((\xi
   +1) M_W^6+\left(4 (D-2) p^2-2 M_Z^2 \xi \right)
   M_W^4 \\&\quad +\left(2 M_Z^2 p^2 \xi -p^4 (4 D+\xi -7)\right)
   M_W^2-(D-1) M_Z^4 p^2\big) M_W^2+2 
   \Big(-(D-1) M_Z^4 C_{HW} p^4 \\&\quad +2 M_W^3 M_Z^2
   \sqrt{M_Z^2-M_W^2} \left(p^2-M_W^2\right) C_{HWB}
   \xi +M_W^3 \sqrt{M_Z^2-M_W^2} (M_W-p)
   (M_W+p) C_{HWB} \big((\xi +1) M_W^2 \\&\quad +p^2 (4 D+\xi
   -7)\big)\Big)\Big)\bigg)+\frac{1}{D-1} \bigg(A_0(M_W^2) \Big(36
   \sqrt[4]{2} (D-2) \sqrt{G_{\mu}}  M_W C_{W} p^4 \\&\quad +2
    M_W \left((3-2 D) \sqrt{M_Z^2-M_W^2}
   C_{HWB} M_W^2+4 (D-1) p^2 C_{HW} M_W+(3-2 D)
   \sqrt{M_Z^2-M_W^2} p^2 C_{HWB}\right) \\&\quad +\sqrt{2}
   G_{\mu} \left((\xi +1) M_W^4-M_H^2
   M_W^2-\left(M_Z^2+2 p^2 (2 (D-3) D+\xi +3)\right)
   M_W^2+p^4 (4 D+\xi -7)\right)\Big)\bigg) \\&\quad -\frac{1}{(D-1) M_Z^4} \bigg(A_0(M_Z^2)
   M_W \Big(36 \sqrt[4]{2} (D-2) \sqrt{G_{\mu}} 
   M_W^2 M_Z^2 \left(M_W^2-M_Z^2-p^2\right)
   C_{W} p^2+\sqrt{2} G_{\mu} M_W \big((\xi +1) M_W^6 \\&\quad +2
   \left(2 (D-2) \left(M_Z^2+p^2\right)-M_Z^2 \xi \right)
   M_W^4+\left((9-4 D) M_Z^4+2 p^2 \left(2 (D-2)^2+\xi \right)
   M_Z^2-p^4 (4 D+\xi -7)\right) M_W^2 \\&\quad -M_Z^4
   \left(M_Z^2+p^2\right)\big)+2 
   \sqrt{M_Z^2-M_W^2} C_{HWB} \big((\xi +1)
   M_W^6+\left((2 D-\xi -5) M_Z^2+4 (D-2) p^2\right)
   M_W^4 \\&\quad +\left(-2 (D-2) M_Z^4+p^2 (2 D (2 D-7)+\xi +13)
   M_Z^2-p^4 (4 D+\xi -7)\right) M_W^2-2 (D-1) M_Z^4
   p^2\big)\Big)\bigg) \\&\quad +\frac{1}{(D-1) M_Z^4} \bigg( B_0(p^2, M_Z^2, M_W^2 \xi) M_W
   (M_W-p) (M_W+p) \Big(-\sqrt{2} G_{\mu} M_W
   \left(M_W^2-2 M_Z^2+p^2\right) \big(M_Z^4+ \\&\quad \left((4
   D-6) p^2-2 M_W^2 \xi \right) M_Z^2+\left(p^2-M_W^2 \xi
   \right)^2\big)-2  \sqrt{M_Z^2-M_W^2}
   C_{HWB} \big(\xi ^2 M_W^6-\xi  \left(M_Z^2 (\xi
   +2)-p^2 (\xi -2)\right) M_W^4 \\&\quad +\left((2 \xi +1) M_Z^4+2 p^2
   (-\xi +D (\xi +2)-3) M_Z^2+p^4 (1-2 \xi )\right)
   M_W^2-M_Z^6+p^6+(2 D-5) M_Z^2 p^4 \\&\quad +3 (3-2 D) M_Z^4
   p^2\big)\Big) \bigg) \Bigg) \numberthis
\end{align*}
which yields the mass shift
\begin{align*}
\delta M_W^2 &= \Pi_{WW}^T(M_W^2) \\
&= \frac{1}{16 \pi ^2} \Bigg(\frac{A_0(M_H^2) \left(\sqrt{2} G_{\mu} \left(M_H^2-(D-1)
   M_W^2\right)-10 (D-1)  M_W^2
   C_{HW}\right)}{D-1} \\&\quad -\frac{6 \sqrt{2} A_0(M_T^2) G_{\mu}
   \left(M_T^2-(D-2) M_W^2\right)}{D-1}-\frac{1}{D-1} \bigg(A_0(M_W^2)
   \bigg(-36 \sqrt[4]{2} (D-2) \sqrt{G_{\mu}}  M_W^3
   C_{W} \\&\quad +\sqrt{2} G_{\mu} \left(4 (D-3) (D-1)
   M_W^2+M_H^2+M_Z^2\right)+4  M_W
   \left((2 D-3) C_{HWB} \sqrt{M_Z^2-M_W^2}-2 (D-1)
   M_W C_{HW}\right)\bigg)\bigg) \\&\quad -2 A_0(M_W^2 \xi) M_W^2
   \left(\sqrt{2} G_{\mu}+2 
   C_{HW}\right)+\frac{1}{(D-1) M_Z^2} \bigg(A_0(M_Z^2) \Big(36 \sqrt[4]{2} (D-2)
   \sqrt{G_{\mu}}  M_W^3 M_Z^2 C_{W} \\&\quad +\sqrt{2}
   G_{\mu} \left(-4 (D-2) (D-1) M_W^4+4 (D-2) M_W^2
   M_Z^2+M_Z^4\right)-4  M_W C_{HWB}
   \sqrt{M_Z^2-M_W^2} \big(2 (D-2) (D-1) M_W^2 \\&\quad +(3-2 D)
   M_Z^2\big)\Big) \bigg) -A_0(M_Z^2 \xi)
   M_W^2 \left(\sqrt{2} G_{\mu}+2 
   C_{HW}\right)-\frac{18 \sqrt{2} B_0(M_W^2, 0, 0) (D-2) G_{\mu}
   M_W^4}{D-1} \\&\quad +\frac{6 \sqrt{2} B_0(M_W^2, 0, M_T^2) G_{\mu}
   (M_T-M_W) (M_T+M_W) \left((D-2)
   M_W^2+M_T^2\right)}{D-1}+\frac{1}{M_Z^2} \bigg(16 B_0(M_W^2, 0, M_W^2) M_W^4
   \\&\quad \times \left(\sqrt{2} G_{\mu} \left(M_Z^2-M_W^2\right)-2
    M_W C_{HWB}
   \sqrt{M_Z^2-M_W^2}\right) \bigg) -\frac{1}{D-1} \bigg(B_0(M_W^2, M_H^2, M_W^2)
   \Big(\sqrt{2} G_{\mu} \big(4 (D-1) M_W^4 \\&\quad +M_H^4-4
   M_H^2 M_W^2\big)-8 (D-1)  M_W^2
   C_{HW} \left(M_H^2-2
   M_W^2\right)\Big)\bigg)+\frac{1}{(D-1) M_Z^2} \bigg(B_0(M_W^2, M_W^2, M_Z^2) \left(4
   M_W^2-M_Z^2\right) \\&\quad \times \Big(36 \sqrt[4]{2} (D-2)
   \sqrt{G_{\mu}}  M_W^3 M_Z^2 C_{W}+\sqrt{2}
   G_{\mu} \left(4 (D-1) M_W^4+4 (2 D-3) M_W^2
   M_Z^2+M_Z^4\right) \\&\quad +4  M_W C_{HWB}
   \sqrt{M_Z^2-M_W^2} \left(2 (D-1) M_W^2+(2 D-3)
   M_Z^2\right)\Big)\bigg)\Bigg) \numberthis
\end{align*}

The transverse $Z$ two-point function is
\begin{align*}
\Pi_{ZZ}^T(p^2) &= \frac{1}{48
   \pi ^2} \Bigg(-\frac{1}{(D-1)
   M_Z^2} \bigg(B_0(p^2, 0, 0) (D-2) \Big(\sqrt{2} G_{\mu} \left(160
   M_W^4-200 M_Z^2 M_W^2+103 M_Z^4\right) \\&\quad +40
    M_W \left(8 M_W^2-5 M_Z^2\right)
   \sqrt{M_Z^2-M_W^2} C_{HWB}\Big) p^2\bigg)+\frac{1}{(D-1)
   M_Z^2} \bigg(2 A_0(M_T^2) (D-2) \Big(\sqrt{2} G_{\mu} \big(32
   M_W^4 \\&\quad -40 M_Z^2 M_W^2+17 M_Z^4\big)+8
    M_W \left(8 M_W^2-5 M_Z^2\right)
   \sqrt{M_Z^2-M_W^2} C_{HWB}\Big)\bigg) -\frac{1}{(1-D)
   M_Z^2} \bigg(B_0(p^2, M_T^2, M_T^2) \\&\quad \times \Big(\sqrt{2} G_{\mu} \left(2
   M_T^2 \left(-64 M_W^4+80 M_Z^2 M_W^2+(9 D-43)
   M_Z^4\right)-(D-2) \left(32 M_W^4-40 M_Z^2
   M_W^2+17 M_Z^4\right) p^2\right) \\&\quad -8  M_W
   \left(8 M_W^2-5 M_Z^2\right) \sqrt{M_Z^2-M_W^2}
   \left(4 M_T^2+(D-2) p^2\right) C_{HWB}\Big)\bigg)-\frac{1}{M_Z^2} \bigg(3 A_0(M_Z^2 \xi) \Big(\sqrt{2} G_{\mu}
   M_Z^4+2  p^2 \\&\quad \times \Big((C_{HW}-C_{HB})
   M_W^2+\sqrt{M_Z^2-M_W^2} C_{HWB}
   M_W+M_Z^2 C_{HB}\Big)\Big)\bigg)+\frac{1}{(D-1)
   M_Z^2} \bigg(3
   B_0(p^2, M_W^2, M_W^2) \\&\quad \times \left(4 M_W^2-p^2\right) \Big(36 \sqrt[4]{2} (D-2)
   \sqrt{G_{\mu}}  p^2 C_{W} M_W^3+4 
   \sqrt{M_Z^2-M_W^2} \left(2 (D-1) M_W^2+(2 D-3)
   p^2\right) C_{HWB} M_W \\&\quad +\sqrt{2} G_{\mu} \left(4 (D-1)
   M_W^4+4 (2 D-3) p^2 M_W^2+p^4\right)\Big)\bigg)+\frac{3 \sqrt{2} B_0(p^2, M_W^2 \xi, M_W^2 \xi) G_{\mu}
   \left(M_Z^4-p^4\right) \left(p^2-4 M_W^2 \xi \right)}{(D-1)
   M_Z^2} \\&\quad +\frac{1}{(D-1) p^2} \bigg(3 A_0(M_Z^2) \Big(\sqrt{2} G_{\mu}
   \left(-M_H^2+M_Z^2+p^2\right) M_Z^2+8 (D-1) 
   p^2 \Big((C_{HW}-C_{HB})
   M_W^2 \\&\quad +\sqrt{M_Z^2-M_W^2} C_{HWB}
   M_W+M_Z^2 C_{HB}\Big)\Big)\bigg)-\frac{1}{(D-1) p^2} \bigg(3
   B_0(p^2, M_H^2, M_Z^2) \Big(\sqrt{2} G_{\mu} \big(p^4 \\&\quad -2 \left(M_H^2+(3-2
   D) M_Z^2\right) p^2+\left(M_H^2-M_Z^2\right)^2\big)
   M_Z^2+8 (D-1)  p^2
   \left(-M_H^2+M_Z^2+p^2\right)
   \Big((C_{HW}-C_{HB})
   M_W^2 \\&\quad +\sqrt{M_Z^2-M_W^2} C_{HWB}
   M_W+M_Z^2 C_{HB}\Big)\Big)\bigg)+\frac{1}{(D-1) M_Z^2
   p^2} \bigg(3
   A_0(M_H^2) \Big(\sqrt{2} G_{\mu} M_Z^4
   \left(M_H^2-M_Z^2-(D-2) p^2\right) \\&\quad -2 (D-1)  p^2
   \left(4 M_Z^2+p^2\right) \left((C_{HW}-C_{HB})
   M_W^2+\sqrt{M_Z^2-M_W^2} C_{HWB}
   M_W+M_Z^2 C_{HB}\right)\Big)\bigg) \\&\quad +\frac{1}{(D-1) M_Z^2
   p^2} \bigg(6 B_0(p^2, M_W^2, M_W^2 \xi) \Big(\sqrt{2} G_{\mu}
   \left(-M_Z^4+p^4+2 M_W^2 (M_Z-p) (M_Z+p)\right)
   \big((\xi -1)^2 M_W^4 \\&\quad +2 p^2 (2 D-\xi -3) M_W^2+p^4\big)+2
    M_W \sqrt{M_Z^2-M_W^2} (M_Z-p)
   (M_Z+p) C_{HWB} \big((\xi -1)^2 M_W^4 \\&\quad +2 (D-2) p^2
   (\xi +1) M_W^2+(3-2 D) p^4\big)\Big)\bigg)-\frac{1}{(D-1) M_Z^2 p^2} \bigg(6 A_0(M_W^2 \xi) \Big(2  \Big((D-1) M_Z^2
   C_{HB} p^4 \\&\quad -(D-1) M_W^2 (C_{HB}-C_{HW})
   p^4+M_W \sqrt{M_Z^2-M_W^2} \left(M_Z^2+(D-2)
   p^2\right) C_{HWB} p^2 \\&\quad +M_W^3 \sqrt{M_Z^2-M_W^2}
   (M_Z-p) (M_Z+p) C_{HWB} (\xi -1)\Big)+\sqrt{2}
   G_{\mu} \big(2 (M_Z-p) (M_Z+p) (\xi -1)
   M_W^4 \\&\quad -(M_Z-p) (M_Z+p) \left((\xi -1) M_Z^2+p^2
   (4 D+\xi -7)\right) M_W^2+(D-1) M_Z^4
   p^2\big)\Big)\bigg) \\&\quad +\frac{1}{(D-1) M_Z^2 p^2} \bigg(6 A_0(M_W^2) \Big(36
   \sqrt[4]{2} (D-2) \sqrt{G_{\mu}}  M_W^3 C_{W}
   p^4+\sqrt{2} G_{\mu} \big(2 \big(M_Z^2 (\xi -1) \\&\quad -p^2 (2 (D-3)
   D+\xi +3)\big) M_W^4+\left(-(\xi -1) M_Z^4-2 p^2
   M_Z^2+p^4 (4 D+\xi -7)\right) M_W^2+M_Z^4 p^2\big) \numberthis \\&\quad +2
    M_W \sqrt{M_Z^2-M_W^2} C_{HWB}
   \left(\left(M_Z^2 (\xi -1)-p^2 (4 (D-3) D+\xi +7)\right)
   M_W^2+(2 D-3) p^2
   \left(M_Z^2+p^2\right)\right)\Big)\bigg)\Bigg)
\end{align*} 
which yields the mass shift
\begin{align*}
\delta M_Z^2 &= \Pi_{ZZ}^T(M_Z^2) \\
&= -\frac{1}{48
   \pi ^2} \Bigg(-\frac{1}{D-1} \bigg(3 A_0(M_H^2) \Big(\sqrt{2} G_{\mu} \left(M_H^2-(D-1)
   M_Z^2\right)-10 (D-1)  \Big(M_W C_{HWB}
   \sqrt{M_Z^2-M_W^2} \\&\quad +M_W^2
   (C_{HW}-C_{HB})+M_Z^2
   C_{HB}\Big)\Big) \bigg)-\frac{1}{(D-1) M_Z^2} \bigg(2 A_0(M_T^2) (D-2) \Big(\sqrt{2}
   G_{\mu} \left(32 M_W^4-40 M_W^2 M_Z^2+17
   M_Z^4\right) \\&\quad +8  M_W C_{HWB} \left(8
   M_W^2-5 M_Z^2\right)
   \sqrt{M_Z^2-M_W^2}\Big)\bigg)-\frac{1}{(D-1)
   M_Z^2} \bigg(6
   A_0(M_W^2) \Big(36 \sqrt[4]{2} (D-2) \sqrt{G_{\mu}} 
   M_W^3 M_Z^2 C_{W} \\&\quad +\sqrt{2} G_{\mu} \left(-4 (D-2)
   (D-1) M_W^4+4 (D-2) M_W^2 M_Z^2+M_Z^4\right)-4
    M_W C_{HWB} \sqrt{M_Z^2-M_W^2}
   \big(2 (D-2) (D-1) M_W^2 \\&\quad +(3-2 D) M_Z^2\big)\Big)\bigg)+6 A_0(M_W^2 \xi) \left(\sqrt{2} G_{\mu} M_Z^2+2
    \left(M_W C_{HWB}
   \sqrt{M_Z^2-M_W^2}+M_W^2
   (C_{HW}-C_{HB})+M_Z^2
   C_{HB}\right)\right) \\&\quad +\frac{3 A_0(M_Z^2) \left(\sqrt{2} G_{\mu}
   \left(M_H^2-2 M_Z^2\right)-8 (D-1) 
   \left(M_W C_{HWB} \sqrt{M_Z^2-M_W^2}+M_W^2
   (C_{HW}-C_{HB})+M_Z^2
   C_{HB}\right)\right)}{D-1} \\&\quad +3 A_0(M_Z^2 \xi) \left(\sqrt{2}
   G_{\mu} M_Z^2+2  \left(M_W C_{HWB}
   \sqrt{M_Z^2-M_W^2}+M_W^2
   (C_{HW}-C_{HB})+M_Z^2
   C_{HB}\right)\right) \\&\quad +\frac{B_0(M_Z^2, 0, 0) (D-2) \left(\sqrt{2}
   G_{\mu} \left(160 M_W^4-200 M_W^2 M_Z^2+103
   M_Z^4\right)+40  M_W C_{HWB} \left(8
   M_W^2-5 M_Z^2\right)
   \sqrt{M_Z^2-M_W^2}\right)}{D-1} \\&\quad +\frac{1}{D-1} \bigg(3 B_0(M_Z^2, M_H^2, M_Z^2)
   \Big(\sqrt{2} G_{\mu} \left(4 (D-1) M_Z^4+M_H^4-4
   M_H^2 M_Z^2\right)-8 (D-1)  \left(M_H^2-2
   M_Z^2\right) \\&\quad \times \left(M_W C_{HWB}
   \sqrt{M_Z^2-M_W^2}+M_W^2
   (C_{HW}-C_{HB})+M_Z^2
   C_{HB}\right)\Big)\bigg)+\frac{1}{(D-1) M_Z^2} \bigg(B_0(M_Z^2, M_T^2, M_T^2)  \\&\quad \times \Big(\sqrt{2}
   G_{\mu} \left(2 M_T^2 \left((43-9 D) M_Z^4+64
   M_W^4-80 M_W^2 M_Z^2\right)+(D-2) M_Z^2 \left(32
   M_W^4-40 M_W^2 M_Z^2+17 M_Z^4\right)\right) \\&\quad +8
    M_W C_{HWB} \left(8 M_W^2-5
   M_Z^2\right) \sqrt{M_Z^2-M_W^2} \left((D-2)
   M_Z^2+4 M_T^2\right)\Big)\bigg)+\frac{1}{(D-1) M_Z^2} \bigg(3
   B_0(M_Z^2, M_W^2, M_W^2) \\&\quad \times \left(M_Z^2-4 M_W^2\right) \Big(36 \sqrt[4]{2}
   (D-2) \sqrt{G_{\mu}}  M_W^3 M_Z^2
   C_{W}+\sqrt{2} G_{\mu} \left(4 (D-1) M_W^4+4 (2 D-3)
   M_W^2 M_Z^2+M_Z^4\right) \\&\quad +4  M_W
   C_{HWB} \sqrt{M_Z^2-M_W^2} \left(2 (D-1)
   M_W^2+(2 D-3) M_Z^2\right)\Big)\bigg) \Bigg) \numberthis
\end{align*}
and the wave function renormalization
\begin{align*}
\delta Z_Z &= -\frac{\partial \Pi_{ZZ}^T(p^2)}{p^2}\bigg|_{p^2 = M_Z^2} \\
&= \frac{1}{48 \pi ^2} \Bigg( \frac{3 \sqrt{2} A_0(M_Z^2) G_{\mu}
   \left(M_Z^2-M_H^2\right)}{(D-1) M_Z^2}+\frac{6
   A_0(M_Z^2 \xi)  \left((C_{HW}-C_{HB})
   M_W^2+\sqrt{M_Z^2-M_W^2} C_{HWB}
   M_W+M_Z^2
   C_{HB}\right)}{M_Z^2} \\&\quad +\frac{B_0(M_Z^2, M_T^2, M_T^2) (D-2)
   \left(\sqrt{2} G_{\mu} \left(32 M_W^4-40 M_Z^2
   M_W^2+17 M_Z^4\right)+8  M_W \left(8
   M_W^2-5 M_Z^2\right) \sqrt{M_Z^2-M_W^2}
   C_{HWB}\right)}{(D-1) M_Z^2} \\&\quad +\frac{(D-2) B'_0(M_Z^2, 0, 0)
   \left(\sqrt{2} G_{\mu} \left(160 M_W^4-200 M_Z^2
   M_W^2+103 M_Z^4\right)+40  M_W \left(8
   M_W^2-5 M_Z^2\right) \sqrt{M_Z^2-M_W^2}
   C_{HWB}\right)}{D-1} \\&\quad +\frac{B_0(M_Z^2, 0, 0) (D-2) \left(\sqrt{2}
   G_{\mu} \left(160 M_W^4-200 M_Z^2 M_W^2+103
   M_Z^4\right)+40  M_W \left(8 M_W^2-5
   M_Z^2\right) \sqrt{M_Z^2-M_W^2}
   C_{HWB}\right)}{(D-1) M_Z^2} \\&\quad +\frac{1}{(D-1)
   M_Z^2} \bigg(B'_0(M_Z^2, M_T^2, M_T^2)
   \Big(\sqrt{2} G_{\mu} \Big(2 \left(64 M_W^4-80 M_Z^2
   M_W^2+(43-9 D) M_Z^4\right) M_T^2+(D-2) M_Z^2
   \big(32 M_W^4 \\&\quad -40 M_Z^2 M_W^2+17
   M_Z^4\big)\Big)+8  M_W \left(8 M_W^2-5
   M_Z^2\right) \sqrt{M_Z^2-M_W^2} \left(4
   M_T^2+(D-2) M_Z^2\right) C_{HWB}\Big)\bigg) \\&\quad +\frac{1}{(D-1)
   M_Z^2} \bigg(3 A_0(M_H^2) \Big(\sqrt{2} G_{\mu}
   (M_H-M_Z) (M_H+M_Z)+2 (D-1) 
   \Big((C_{HW}-C_{HB})
   M_W^2 \\&\quad +\sqrt{M_Z^2-M_W^2} C_{HWB}
   M_W+M_Z^2 C_{HB}\Big)\Big) \bigg) +\frac{1}{(D-1)
   M_Z^2} \bigg(3 B_0(M_Z^2, M_H^2, M_Z^2) \Big(8 (D-1)  M_Z^2
   \Big((C_{HW}-C_{HB})
   M_W^2 \\&\quad +\sqrt{M_Z^2-M_W^2} C_{HWB}
   M_W+M_Z^2 C_{HB}\Big)-\sqrt{2} G_{\mu} M_H^2
   \left(M_H^2-2 M_Z^2\right)\Big)\bigg) +\frac{1}{D-1} \bigg(3 B'_0(M_Z^2, M_H^2, M_Z^2)  \\&\quad \times \Big(\sqrt{2} G_{\mu}
   \left(M_H^4-4 M_Z^2 M_H^2+4 (D-1) M_Z^4\right)-8
   (D-1)  \left(M_H^2-2 M_Z^2\right)
   \Big((C_{HW}-C_{HB})
   M_W^2 \\&\quad +\sqrt{M_Z^2-M_W^2} C_{HWB}
   M_W+M_Z^2 C_{HB}\Big)\Big)\bigg)+\frac{1}{(D-1) M_Z^2} \bigg(3
   B'_0(M_Z^2, M_W^2, M_W^2) \left(M_Z^2-4 M_W^2\right) \\&\quad \times \Big(36 \sqrt[4]{2}
   (D-2) \sqrt{G_{\mu}}  M_Z^2 C_{W} M_W^3+4
    \sqrt{M_Z^2-M_W^2} \left(2 (D-1) M_W^2+(2
   D-3) M_Z^2\right) C_{HWB} M_W \\&\quad +\sqrt{2} G_{\mu}
   \left(4 (D-1) M_W^4+4 (2 D-3) M_Z^2
   M_W^2+M_Z^4\right)\Big)\bigg)+\frac{1}{(D-1) M_Z^2} \bigg(3
   B_0(M_Z^2, M_W^2, M_W^2) \\&\quad \times \Big(-72 \sqrt[4]{2} (D-2) \sqrt{G_{\mu}} 
   \left(2 M_W^2-M_Z^2\right) C_{W} M_W^3-8
    \sqrt{M_Z^2-M_W^2} \left((3 D-5) M_W^2+(3-2
   D) M_Z^2\right) C_{HWB} M_W \\&\quad +\sqrt{2} G_{\mu}
   \left((44-28 D) M_W^4+16 (D-2) M_Z^2 M_W^2+3
   M_Z^4\right)\Big)\bigg) +\frac{6 \sqrt{2}
   B_0(M_Z^2, M_W^2 \xi, M_W^2 \xi) G_{\mu} \left(M_Z^2-4 M_W^2 \xi
   \right)}{D-1} \\&\quad +\frac{1}{(D-1) M_Z^4} \bigg(12 B_0(M_Z^2, M_W^2, M_W^2 \xi) \Big(\sqrt{2} G_{\mu}
   (M_W-M_Z) (M_W+M_Z) \big((\xi -1)^2
   M_W^4+2 M_Z^2 (2 D-\xi -3)
   M_W^2 \\&\quad +M_Z^4\big)+ M_W
   \sqrt{M_Z^2-M_W^2} C_{HWB} \left((\xi -1)^2
   M_W^4+2 (D-2) M_Z^2 (\xi +1) M_W^2+(3-2 D)
   M_Z^4\right)\Big)\bigg) \\&\quad +\frac{1}{(D-1)
   M_Z^4} \bigg(12 A_0(M_W^2)
   M_W \Big(-18 \sqrt[4]{2} (D-2) \sqrt{G_{\mu}} 
   M_W^2 C_{W} M_Z^2+
   \sqrt{M_Z^2-M_W^2} C_{HWB} \\&\quad \times \left((\xi -1)
   M_W^2+(3-2 D) M_Z^2\right)+\sqrt{2} G_{\mu} M_W
   \left(M_W^2 (\xi -1)-M_Z^2 (2 D+\xi -4)\right)\Big)\bigg) \\&\quad + \frac{1}{(D-1) M_Z^4} \bigg(12 A_0(M_W^2 \xi) \Big(\sqrt{2} G_{\mu}
   \left(M_Z^2 (2 D+\xi -4)-M_W^2 (\xi -1)\right)
   M_W^2+ \Big((D-1) C_{HB} M_Z^4 \numberthis \\&\quad -(D-1)
   M_W^2 (C_{HB}-C_{HW}) M_Z^2+(D-2) M_W
   \sqrt{M_Z^2-M_W^2} C_{HWB} M_Z^2-M_W^3
   \sqrt{M_Z^2-M_W^2} C_{HWB} (\xi
   -1)\Big)\Big)\bigg) \Bigg)
\end{align*}

The $\gamma - Z$ two-point function is
\begin{align*}
\Pi_{\gamma Z}^T(p^2) &= \frac{1}{48 \pi ^2} \Bigg(\frac{3 \sqrt{2} B_0(p^2, M_W^2 \xi, M_W^2 \xi) G_{\mu}
   \sqrt{M_Z^2-M_W^2} \left(4 M_W^2 \xi -p^2\right)
   p^4}{(D-1) M_W M_Z^2} \\&\quad +\frac{3 A_0(M_Z^2 \xi) 
   \left(2 C_{HWB} M_W^2+2 \sqrt{M_Z^2-M_W^2}
   (C_{HB}-C_{HW}) M_W-M_Z^2 C_{HWB}\right)
   p^2}{M_Z^2} \\&\quad +\frac{20 B_0(p^2, 0, 0) (D-2) M_W \left(\sqrt{2}
   G_{\mu} \sqrt{M_Z^2-M_W^2} \left(5 M_Z^2-8
   M_W^2\right)+ M_W \left(16 M_W^2-13
   M_Z^2\right) C_{HWB}\right) p^2}{(D-1) M_Z^2} \\&\quad +6
   B_0(p^2, M_H^2, M_Z^2)  \left(-M_H^2+M_Z^2+p^2\right)
   \left(2 C_{HWB} M_W^2+2 \sqrt{M_Z^2-M_W^2}
   (C_{HB}-C_{HW}) M_W-M_Z^2 C_{HWB}\right) \\&\quad +6
   A_0(M_Z^2)  \left(-2 C_{HWB} M_W^2+2
   \sqrt{M_Z^2-M_W^2} (C_{HW}-C_{HB})
   M_W+M_Z^2 C_{HWB}\right) \\&\quad -\frac{3 A_0(M_H^2) 
   \left(2 M_Z^2+p^2\right) \left(-2 C_{HWB} M_W^2+2
   \sqrt{M_Z^2-M_W^2} (C_{HW}-C_{HB})
   M_W+M_Z^2 C_{HWB}\right)}{M_Z^2} \\&\quad +\frac{1}{(D-1)
   M_Z^2} \bigg(4
   B_0(p^2, M_T^2, M_T^2) M_W \left(4 M_T^2+(D-2) p^2\right)
   \Big(\sqrt{2} G_{\mu} \sqrt{M_Z^2-M_W^2} \left(5
   M_Z^2-8 M_W^2\right) \\&\quad + M_W \left(16
   M_W^2-13 M_Z^2\right) C_{HWB}\Big)\bigg) +\frac{1}{(D-1) M_Z^2} \bigg(8 A_0(M_T^2) (D-2) M_W \Big(\sqrt{2} G_{\mu}
   \sqrt{M_Z^2-M_W^2} \left(8 M_W^2-5
   M_Z^2\right) \\&\quad + M_W \left(13 M_Z^2-16
   M_W^2\right) C_{HWB}\Big)\bigg)  -\frac{1}{(D-1) M_W
   M_Z^2} \bigg(3
   B_0(p^2, M_W^2, M_W^2) \left(4 M_W^2-p^2\right) \\&\quad \times \Big(-36 \sqrt[4]{2} (D-2)
   \sqrt{G_{\mu}}  \sqrt{M_Z^2-M_W^2} p^2 C_{W}
   M_W^3+2  \left(2 M_W^2-M_Z^2\right) \left(2
   (D-1) M_W^2+(2 D-3) p^2\right) C_{HWB} M_W \\&\quad -\sqrt{2}
   G_{\mu} \sqrt{M_Z^2-M_W^2} \left(4 (D-1) M_W^4+4 (2
   D-3) p^2 M_W^2+p^4\right)\Big)\bigg)+\frac{1}{(D-1) M_W M_Z^2 p^2} \bigg(6 B_0(p^2, M_W^2, M_W^2 \xi) \\&\quad \times \Big(\sqrt{2} G_{\mu}
   \sqrt{M_Z^2-M_W^2} \left(p^4+M_W^2 \left(M_Z^2-2
   p^2\right)\right) \left((\xi -1)^2 M_W^4+2 p^2 (2 D-\xi -3)
   M_W^2+p^4\right) \\&\quad - M_W \left(\left(M_Z^2-2
   p^2\right) M_W^2+M_Z^2 p^2\right) C_{HWB} \left((\xi
   -1)^2 M_W^4+2 (D-2) p^2 (\xi +1) M_W^2+(3-2 D)
   p^4\right)\Big)\bigg) \\&\quad -\frac{1}{(D-1) M_Z^2 p^2} \bigg(6
   A_0(M_W^2 \xi) \Big( \Big(-\left(M_Z^2-2 p^2\right)
   C_{HWB} (\xi -1) M_W^4-p^2 C_{HWB} \left(\xi 
   M_Z^2+2 (D-2) p^2\right) M_W^2 \\&\quad -2 (D-1)
   \sqrt{M_Z^2-M_W^2} p^4 (C_{HB}-C_{HW})
   M_W+(D-2) M_Z^2 p^4 C_{HWB}\Big) \\&\quad +\sqrt{2} G_{\mu}
   M_W \sqrt{M_Z^2-M_W^2} \left(\left(M_Z^2-2
   p^2\right) (\xi -1) M_W^2+p^2 \left((3-2 D) M_Z^2+p^2 (4
   D+\xi -7)\right)\right)\Big)\bigg) \\&\quad +\frac{1}{(D-1) M_Z^2 p^2} \bigg(6
   A_0(M_W^2) \Big(36 \sqrt[4]{2} (D-2) \sqrt{G_{\mu}} 
   M_W^2 \sqrt{M_Z^2-M_W^2} C_{W} p^4+\sqrt{2}
   G_{\mu} M_W \sqrt{M_Z^2-M_W^2} \\&\quad \times \left((4 D+\xi -7)
   p^4-M_Z^2 p^2+M_W^2 \left(M_Z^2 (\xi -1)-2 p^2 (2
   (D-3) D+\xi +3)\right)\right)+ C_{HWB} \big(\big(2 p^2
   (4 (D-3) D+\xi +7) \\&\quad -M_Z^2 (\xi -1)\big) M_W^4-p^2 \left((2
   D (2 D-5)+\xi +4) M_Z^2+2 (2 D-3) p^2\right) M_W^2+(2 D-3)
   M_Z^2 p^4\big)\Big)\bigg)\Bigg) \numberthis
\end{align*}
which yields the on-shell mixing
\begin{align*}
\Pi_{\gamma Z}(M_Z^2) &= \frac{1}{48 \pi ^2} \Bigg(-\frac{3 \sqrt{2} B_0(M_Z^2, M_W^2 \xi, M_W^2 \xi) G_{\mu}
   \sqrt{M_Z^2-M_W^2} \left(M_Z^2-4 M_W^2 \xi
   \right) M_Z^2}{(D-1) M_W}+9 A_0(M_H^2) \\&\quad \times \left(2
   C_{HWB} M_W^2+2 \sqrt{M_Z^2-M_W^2}
   (C_{HB}-C_{HW}) M_W-M_Z^2 C_{HWB}\right)+3
   A_0(M_Z^2 \xi)  \Big(2 C_{HWB} M_W^2 \\&\quad +2
   \sqrt{M_Z^2-M_W^2} (C_{HB}-C_{HW})
   M_W-M_Z^2 C_{HWB}\Big)+6 A_0(M_Z^2) 
   \Big(-2 C_{HWB} M_W^2 \\&\quad +2 \sqrt{M_Z^2-M_W^2}
   (C_{HW}-C_{HB}) M_W+M_Z^2 C_{HWB}\Big)+6
   B_0(M_Z^2, M_H^2, M_Z^2)  \left(M_H^2-2 M_Z^2\right) \Big(-2
   C_{HWB} M_W^2 \\&\quad +2 \sqrt{M_Z^2-M_W^2}
   (C_{HW}-C_{HB}) M_W+M_Z^2
   C_{HWB}\Big)+\frac{1}{D-1} \bigg(20 B_0(M_Z^2, 0, 0) (D-2) M_W \\&\quad \times
   \left(\sqrt{2} G_{\mu} \sqrt{M_Z^2-M_W^2} \left(5
   M_Z^2-8 M_W^2\right)+ M_W \left(16
   M_W^2-13 M_Z^2\right) C_{HWB}\right)\bigg) \\&\quad +\frac{1}{(D-1)
   M_Z^2} \bigg(4
   B_0(M_Z^2, M_T^2, M_T^2) M_W \left(4 M_T^2+(D-2) M_Z^2\right)
   \Big(\sqrt{2} G_{\mu} \sqrt{M_Z^2-M_W^2} \left(5
   M_Z^2-8 M_W^2\right) \\&\quad + M_W \left(16
   M_W^2-13 M_Z^2\right) C_{HWB}\Big)\bigg) +\frac{1}{(D-1) M_Z^2} \bigg(8 A_0(M_T^2) (D-2) M_W \Big(\sqrt{2} G_{\mu}
   \sqrt{M_Z^2-M_W^2} \left(8 M_W^2-5
   M_Z^2\right) \\&\quad + M_W \left(13 M_Z^2-16
   M_W^2\right) C_{HWB}\Big)\bigg)+\frac{1}{(D-1) M_W
   M_Z^2} \bigg(3
   B_0(M_Z^2, M_W^2, M_W^2) \left(4 M_W^2-M_Z^2\right) \\&\quad \times \Big(36 \sqrt[4]{2}
   (D-2) \sqrt{G_{\mu}}  M_Z^2
   \sqrt{M_Z^2-M_W^2} C_{W} M_W^3-2 
   \left(2 M_W^2-M_Z^2\right) \left(2 (D-1) M_W^2+(2 D-3)
   M_Z^2\right) C_{HWB} M_W \\&\quad +\sqrt{2} G_{\mu}
   \sqrt{M_Z^2-M_W^2} \left(4 (D-1) M_W^4+4 (2 D-3)
   M_Z^2 M_W^2+M_Z^4\right)\Big)\bigg) \\&\quad +\frac{1}{(D-1)
   M_W M_Z^2} \bigg( 6 B_0(M_Z^2, M_W^2, M_W^2 \xi) (M_W-M_Z)
   (M_W+M_Z) \Big( M_W C_{HWB}
   \big((\xi -1)^2 M_W^4 \\&\quad +2 (D-2) M_Z^2 (\xi +1)
   M_W^2+(3-2 D) M_Z^4\big)-\sqrt{2} G_{\mu}
   \sqrt{M_Z^2-M_W^2} \big((\xi -1)^2 M_W^4+2
   M_Z^2 (2 D-\xi -3) M_W^2 \\&\quad +M_Z^4\big)\Big) \bigg) -\frac{1}{(D-1)
   M_Z^2} \bigg(6 A_0(M_W^2) \Big(-36 \sqrt[4]{2} (D-2)
   \sqrt{G_{\mu}}  M_W^2 \sqrt{M_Z^2-M_W^2}
   C_{W} M_Z^2 \\&\quad +\sqrt{2} G_{\mu} M_W
   \sqrt{M_Z^2-M_W^2} \left(M_W^2 (4 (D-3) D+\xi
   +7)-M_Z^2 (4 D+\xi -8)\right)- C_{HWB} \big((8
   (D-3) D+\xi +15) M_W^4 \\&\quad -M_Z^2 \left(4 D^2-6 D+\xi -2\right)
   M_W^2+(2 D-3) M_Z^4\big)\Big)\bigg) +\frac{1}{(D-1)
   M_Z^2} \bigg(6 A_0(M_W^2 \xi) \numberthis \\&\quad \times \Big(\sqrt{2} G_{\mu} M_W
   \sqrt{M_Z^2-M_W^2} \left(M_W^2 (\xi -1)-M_Z^2 (2
   D+\xi -4)\right)+ \Big((C_{HWB}-C_{HWB} \xi )
   M_W^4 \\&\quad +M_Z^2 C_{HWB} (2 D+\xi -4) M_W^2+2 (D-1)
   M_Z^2 \sqrt{M_Z^2-M_W^2} (C_{HB}-C_{HW})
   M_W-(D-2) M_Z^4 C_{HWB}\Big)\Big)\bigg)\Bigg)
\end{align*}
\section{Vertex functions}
\label{sec:appb}
The one loop amplitude for $Z(p + p') \to f(p) \bar{f}(p')$, the decay of a $Z$ boson to a pair of massless fermions with weak isospin $T_3^f$ and charge $Q_f$, is
\begin{equation}
\mathcal{M}_1 = V \bar{u}(p) \slashed{\epsilon}^*(p + p') v(p')
\end{equation}
where the vertex function is 
\begin{align*}
V &= \frac{\sqrt{G_{\mu}}}{4 \cdot 2^{3/4} M_Z^5 \pi ^2} \Bigg( \frac{\sqrt{2} B_0(M_Z^2, M_W^2 \xi, M_W^2 \xi)
   G_{\mu} T_3^f \left(M_Z^2-4 M_W^2 \xi \right)
   M_Z^6}{D-1}+2 \sqrt{2} C_0(M_Z^2, 0, 0, 0, 0, M_Z^2) (D-8) \big(Q_f
   M_W^2 \\&\quad +M_Z^2 (T_3^f-Q_f)\big)^2 \left(2 G_{\mu} \left(Q_f
   M_W^2+M_Z^2 (T_3^f-Q_f)\right)+3 \sqrt{2} 
   M_W \sqrt{M_Z^2-M_W^2} Q_f C_{HWB}\right)
   M_Z^4 \\&\quad +4 C_0(M_Z^2, 0, 0, 0, 0, M_Z^2 \beta) M_W^2 Q_f^2 \Big(\sqrt{2}
   G_{\mu} (M_W-M_Z) (M_W+M_Z) \left(Q_f
   M_W^2+M_Z^2 (T_3^f-Q_f)\right) \\&\quad + M_W
   \sqrt{M_Z^2-M_W^2} \left(3 Q_f M_W^2+M_Z^2 (2
   T_3^f-3 Q_f)\right) C_{HWB}\Big) (\beta  (-D+2 \beta +8)+2)
   M_Z^4+8 \sqrt{2} B_0(0, 0, M_Z^2) \big(Q_f M_W^2 \\&\quad +M_Z^2
   (T_3^f-Q_f)\big)^2 \left(2 G_{\mu} \left(Q_f M_W^2+M_Z^2
   (T_3^f-Q_f)\right)+3 \sqrt{2}  M_W
   \sqrt{M_Z^2-M_W^2} Q_f C_{HWB}\right) M_Z^2 \\&\quad -4
   \sqrt{2} C_{00}(0, M_Z^2, 0, M_Z^2, 0, 0) (D-2) \left(Q_f M_W^2+M_Z^2
   (T_3^f-Q_f)\right)^2 \Big(2 G_{\mu} \left(Q_f M_W^2+M_Z^2
   (T_3^f-Q_f)\right) \\&\quad +3 \sqrt{2}  M_W
   \sqrt{M_Z^2-M_W^2} Q_f C_{HWB}\Big) M_Z^2+8
   C_0(M_Z^2, 0, 0, 0, 0, M_W^2) \left(2 M_W^6-(D-8) M_Z^2 M_W^4+2
   M_Z^4 M_W^2\right) T_3^{f2} \\&\quad \times \left(\sqrt{2} G_{\mu}
   \left(M_Z^2 (Q_f-T_3^f)-M_W^2 (Q_f-2
   T_3^f)\right)- M_W \sqrt{M_Z^2-M_W^2}
   (Q_f-2 T_3^f) C_{HWB}\right) M_Z^2 \\&\quad + \frac{1}{D-2} \bigg(4 C_0(M_Z^2, 0, 0, M_W^2, M_W^2, 0)
   M_W^5 T_3^f \Big(3 \sqrt[4]{2} (3 D-8) \sqrt{G_{\mu}}
    M_W^2 C_{W} M_Z^2+2 \sqrt{2} (D-2) G_{\mu}
   M_W \left(M_W^2+2 M_Z^2\right) \\&\quad +2 (D-2) 
   \sqrt{M_Z^2-M_W^2} \left(M_W^2+M_Z^2\right)
   C_{HWB}\Big) M_Z^2\bigg)+8 B_0(0, 0, M_W^2) M_W^2
   T_3^f \Big(\frac{6 \sqrt[4]{2} (D-3) \sqrt{G_{\mu}} 
   M_Z^2 C_{W} M_W^3}{D-2} \\&\quad +
   \sqrt{M_Z^2-M_W^2} \left((4 (Q_f-2 T_3^f) T_3^f+2)
   M_W^2+M_Z^2 (4 (Q_f-2 T_3^f) T_3^f+1)\right)
   C_{HWB} M_W+2 \sqrt{2} G_{\mu}
   \left(M_W^2+M_Z^2\right) \\&\quad \times \left((2 (Q_f-2 T_3^f)
   T_3^f+1) M_W^2+2 M_Z^2 T_3^f
   (T_3^f-Q_f)\right)\Big) M_Z^2+\frac{1}{(D-2)
   (D-1)} \bigg(B_0(M_Z^2, M_W^2, M_W^2) T_3^f
   \\&\quad \times \Big(-6 \sqrt[4]{2} (D-1) (3 D-8) \sqrt{G_{\mu}} 
   M_Z^2 \left(2 M_W^2-M_Z^2\right) C_{W}
   M_W^3-2 (D-2)  \sqrt{M_Z^2-M_W^2} \big(4
   (D-1) M_W^4 \\&\quad +2 (D-3) M_Z^2 M_W^2+(3-2 D)
   M_Z^4\big) C_{HWB} M_W-\sqrt{2} (D-2) G_{\mu}
   \left(2 M_W^2-M_Z^2\right) \big(4 (D-1) M_W^4+4 (2
   D-3) M_Z^2 M_W^2 \\&\quad +M_Z^4\big)\Big) M_Z^2\bigg)-16 B_0(0, 0, M_Z^2 \beta) M_W^2 Q_f^2 \Big(\sqrt{2} G_{\mu}
   (M_W-M_Z) (M_W+M_Z) \left(Q_f
   M_W^2+M_Z^2 (T_3^f-Q_f)\right) \\&\quad + M_W
   \sqrt{M_Z^2-M_W^2} \left(3 Q_f M_W^2+M_Z^2 (2
   T_3^f-3 Q_f)\right) C_{HWB}\Big) (\beta +1) M_Z^2-4
   B_0(0, M_Z^2 \beta, M_Z^2 \beta) M_W^2 Q_f^2 \\&\quad \times \Big(\sqrt{2} G_{\mu}
   (M_W-M_Z) (M_W+M_Z) \left(Q_f
   M_W^2+M_Z^2 (T_3^f-Q_f)\right) \\&\quad + M_W
   \sqrt{M_Z^2-M_W^2} \left(3 Q_f M_W^2+M_Z^2 (2
   T_3^f-3 Q_f)\right) C_{HWB}\Big) (\xi -1) M_Z^2 \\&\quad +\frac{1}{D-1} \bigg(2
   B_0(M_Z^2, M_W^2, M_W^2 \xi) T_3^f \Big(\sqrt{2} G_{\mu}
   (M_W-M_Z) (M_W+M_Z) \big((\xi -1)^2
   M_W^4 +2 M_Z^2 (2 D-\xi -3)
   M_W^2+M_Z^4\big) \\&\quad + M_W
   \sqrt{M_Z^2-M_W^2} C_{HWB} \big((\xi -1)^2
   M_W^4+2 (D-2) M_Z^2 (\xi +1) M_W^2 \\&\quad +(3-2 D)
   M_Z^4\big)\Big) M_Z^2\bigg)+\frac{1}{D-1} \bigg(2 A_0(M_W^2 \xi)
   T_3^f \Big( M_W \sqrt{M_Z^2-M_W^2}
   C_{HWB} \big(M_Z^2 \big(-8 (D-1) T_3^{f2}+4 (D-1) Q_f
   T_3^f \\&\quad +2 D-3\big)-M_W^2 (\xi -1)\big)+\sqrt{2} G_{\mu}
   \big(-(\xi -1) M_W^4+M_Z^2 \left(-4 Q_f T_3^f+4 \left(-2
   (D-1) T_3^{f2}+D Q_f T_3^f+D\right)+\xi -6\right) M_W^2 \\&\quad -4
   (D-1) M_Z^4 (Q_f-T_3^f) T_3^f\big)\Big)
   M_Z^2\bigg)+2 A_0(M_W^2) T_3^f \bigg(4 \left(-2
   M_W^2-M_Z^2\right) T_3^f \Big(\sqrt{2} G_{\mu}
   \left((Q_f-2 T_3^f) M_W^2+M_Z^2
   (T_3^f-Q_f)\right) \\&\quad + M_W \sqrt{M_Z^2-M_W^2}
   (Q_f-2 T_3^f) C_{HWB}\Big)+\frac{M_W}{(D-2) (D-1)} \bigg(\Big(-6 \sqrt[4]{2}
   (D-4) (D-1) \sqrt{G_{\mu}}  M_W^2 C_{W}
   M_Z^2 \\&\quad -(D-2)  \sqrt{M_Z^2-M_W^2} C_{HWB}
   \left((4 D-\xi -3) M_W^2+(2 D-3) M_Z^2\right)-\sqrt{2} (D-2)
   G_{\mu} M_W \big((4 D-\xi -3) M_W^2 \\&\quad +M_Z^2 (4 D+\xi
   -6)\big)\Big)\bigg)\bigg) M_Z^2-2 \sqrt{2} A_0(M_Z^2)
   \left(Q_f M_W^2+M_Z^2 (T_3^f-Q_f)\right)^2 \Big(2 G_{\mu}
   \left(Q_f M_W^2+M_Z^2 (T_3^f-Q_f)\right) \\&\quad +3 \sqrt{2}
    M_W \sqrt{M_Z^2-M_W^2} Q_f
   C_{HWB}\Big)+2 \sqrt{2} A_0(M_Z^2 \xi) \left(Q_f
   M_W^2+M_Z^2 (T_3^f-Q_f)\right)^2 \Big(2 G_{\mu} \left(Q_f
   M_W^2+M_Z^2 (T_3^f-Q_f)\right) \\&\quad +3 \sqrt{2} 
   M_W \sqrt{M_Z^2-M_W^2} Q_f C_{HWB}\Big)+8
   A_0(M_Z^2 \beta) M_W^2 Q_f^2 \Big(\sqrt{2} G_{\mu}
   (M_W-M_Z) (M_W+M_Z) \left(Q_f
   M_W^2+M_Z^2 (T_3^f-Q_f)\right) \\&\quad + M_W
   \sqrt{M_Z^2-M_W^2} \left(3 Q_f M_W^2+M_Z^2 (2
   T_3^f-3 Q_f)\right) C_{HWB}\Big)+2 B_0(M_Z^2, 0, 0) \bigg(2
   M_W^2 M_Z^2 \\&\quad \times \Big(\sqrt{2} G_{\mu} (M_W-M_Z)
   (M_W+M_Z) \left(Q_f M_W^2+M_Z^2
   (T_3^f-Q_f)\right) \\&\quad + M_W \sqrt{M_Z^2-M_W^2}
   \left(3 Q_f M_W^2+M_Z^2 (2 T_3^f-3 Q_f)\right)
   C_{HWB}\Big) (-D+2 \beta +7) Q_f^2 \\&\quad +\sqrt{2} (D-6)
   \left((T_3^f-Q_f) M_Z^3+M_W^2 Q_f M_Z\right)^2 \Big(2
   G_{\mu} \left(Q_f M_W^2+M_Z^2 (T_3^f-Q_f)\right) +3 \sqrt{2}
    M_W \sqrt{M_Z^2-M_W^2} Q_f
   C_{HWB}\Big) \\&\quad +4 M_W^2 M_Z^2 \left((D-7) M_Z^2-2
   M_W^2\right) T_3^{f2} \Big(\sqrt{2} G_{\mu} \left((Q_f-2
   T_3^f) M_W^2+M_Z^2 (T_3^f-Q_f)\right) \\&\quad +
   M_W \sqrt{M_Z^2-M_W^2} (Q_f-2 T_3^f)
   C_{HWB}\Big)\bigg)\Bigg) \numberthis
\end{align*}
For the $b_L$, there are also top mass effects, and the vertex function in Feynman gauge is
\begin{align*}
V &= \frac{\sqrt{G_{\mu}}}{216 \cdot 2^{3/4} M_Z^3 \pi ^2} \Bigg(-108 M_Z^2 \bigg(2  \sqrt{M_Z^2-M_W^2} C_{HWB}
   C_0\left(M_Z^2,0,0,M_W^2,M_W^2,M_T^2\right) M_W^2 \\&\quad +2 \sqrt{2} G_{\mu}
   \big(C_0\left(M_Z^2,0,0,M_W^2,M_W^2,M_T^2\right) M_W^2+B_0\left(0,M_T^2,M_W^2\right)-M_Z^2
   C_1\left(0,M_Z^2,0,M_T^2,M_W^2,M_W^2\right) \\&\quad +(D-2) C_{00}\left(0,M_Z^2,0,M_T^2,M_W^2,M_W^2\right)\big)
   M_W+2  \sqrt{M_Z^2-M_W^2} C_{HWB} B_0\left(0,M_T^2,M_W^2\right) \\&\quad +2 D  \sqrt{M_Z^2-M_W^2}
   C_{HWB} C_{00}\left(0,M_Z^2,0,M_T^2,M_W^2,M_W^2\right)-4  \sqrt{M_Z^2-M_W^2} C_{HWB}
   C_{00}\left(0,M_Z^2,0,M_T^2,M_W^2,M_W^2\right) \\&\quad +3 \sqrt[4]{2} \sqrt{G_{\mu}}  M_Z^2 C_{W}
   \big(C_0\left(M_Z^2,0,0,M_W^2,M_W^2,M_T^2\right) M_W^2+B_0\left(0,M_T^2,M_W^2\right) \\&\quad +\left(M_Z^2-2
   M_W^2\right) C_1\left(0,M_Z^2,0,M_T^2,M_W^2,M_W^2\right)+(D-4)
   C_{00}\left(0,M_Z^2,0,M_T^2,M_W^2,M_W^2\right)\big)\bigg) M_W^3 \\&\quad +18 M_Z^2 \bigg(4 \sqrt{2} (D-6) G_{\mu}
   C_0\left(M_Z^2,0,0,M_T^2,M_T^2,M_W^2\right) M_W^4+4 D  \sqrt{M_Z^2-M_W^2} C_{HWB}
   C_0\left(M_Z^2,0,0,M_T^2,M_T^2,M_W^2\right) M_W^3 \\&\quad -24  \sqrt{M_Z^2-M_W^2} C_{HWB}
   C_0\left(M_Z^2,0,0,M_T^2,M_T^2,M_W^2\right) M_W^3 \\&\quad -\sqrt{2} G_{\mu} \left(\left(4 (D-6) M_T^2+(D+2) M_Z^2\right)
   C_0\left(M_Z^2,0,0,M_T^2,M_T^2,M_W^2\right)+8 (D-2)
   C_{00}\left(0,M_Z^2,0,M_W^2,M_T^2,M_T^2\right)\right) M_W^2 \\&\quad -4 D  M_T^2 \sqrt{M_Z^2-M_W^2}
   C_{HWB} C_0\left(M_Z^2,0,0,M_T^2,M_T^2,M_W^2\right) M_W \\&\quad +24  M_T^2 \sqrt{M_Z^2-M_W^2}
   C_{HWB} C_0\left(M_Z^2,0,0,M_T^2,M_T^2,M_W^2\right) M_W \\&\quad -8  M_Z^2 \sqrt{M_Z^2-M_W^2}
   C_{HWB} C_0\left(M_Z^2,0,0,M_T^2,M_T^2,M_W^2\right) M_W \\&\quad -8 D  \sqrt{M_Z^2-M_W^2} C_{HWB}
   C_{00}\left(0,M_Z^2,0,M_W^2,M_T^2,M_T^2\right) M_W+16  \sqrt{M_Z^2-M_W^2} C_{HWB}
   C_{00}\left(0,M_Z^2,0,M_W^2,M_T^2,M_T^2\right) M_W \\&\quad +4 \left(\sqrt{2} G_{\mu} \left(4 M_W^2-M_Z^2\right)+4
    M_W \sqrt{M_Z^2-M_W^2} C_{HWB}\right) B_0\left(0,M_T^2,M_W^2\right)+(D-6) \bigg(\sqrt{2} G_{\mu} \left(4
   M_W^2-M_Z^2\right) \\&\quad +4  M_W \sqrt{M_Z^2-M_W^2} C_{HWB}\bigg)
   B_0\left(M_Z^2,M_T^2,M_T^2\right)+2 \sqrt{2} G_{\mu} M_Z^2 \big(\left(2 (D-3) M_T^2+M_Z^2\right)
   C_0\left(M_Z^2,0,0,M_T^2,M_T^2,M_W^2\right) \\&\quad +(D-2)
   C_{00}\left(0,M_Z^2,0,M_W^2,M_T^2,M_T^2\right)\big)\bigg) M_W^2+4 \bigg(6  \sqrt{M_Z^2-M_W^2}
   C_{HWB} M_W^3 \\&\quad +\sqrt{2} G_{\mu} (M_W-M_Z) (M_W+M_Z) \left(2 M_W^2+M_Z^2\right)\bigg) \big(((D-6) \beta -2)
   C_0\left(M_Z^2,0,0,0,0,M_Z^2 \beta \right) M_Z^2+4 B_0\left(0,0,M_Z^2 \beta \right) \\&\quad +(D-6)
   B_0\left(M_Z^2,0,0\right)-2 (D-2) C_{00}\left(0,M_Z^2,0,M_Z^2 \beta ,0,0\right)\big) M_W^2+108 M_T^2 M_Z^2
   \Big(2 \sqrt{2} G_{\mu} M_W (M_W-M_Z) (M_W+M_Z) \\&\quad + \left(2 M_W^2-M_Z^2\right) \sqrt{M_Z^2-M_W^2}
   C_{HWB}\Big) C_0\left(M_Z^2,0,0,M_W^2,M_W^2,M_T^2\right) M_W+108 M_T^2 M_Z^2 \Big(\sqrt{2} G_{\mu}
   \left(M_Z^2-2 M_W^2\right) \\&\quad -2  M_W \sqrt{M_Z^2-M_W^2} C_{HWB}\Big)
   C_{00}\left(0,M_Z^2,0,M_T^2,M_W^2,M_W^2\right)+18 M_T^2 M_Z^2 \bigg(4 \Big(\sqrt{2} G_{\mu} (M_W-M_Z)
   (M_W+M_Z) \\&\quad + M_W \sqrt{M_Z^2-M_W^2} C_{HWB}\Big)
   B_0\left(M_Z^2,M_T^2,M_T^2\right)+\bigg(\sqrt{2} G_{\mu} \left(\left(M_Z^2-4 M_W^2\right) M_T^2+4 M_W^2
   (M_W-M_Z) (M_W+M_Z)\right) \\&\quad +4  M_W \left(M_W^2-M_T^2\right) \sqrt{M_Z^2-M_W^2} C_{HWB}\bigg)
   C_0\left(M_Z^2,0,0,M_T^2,M_T^2,M_W^2\right)-8 \bigg(\sqrt{2} G_{\mu} (M_W-M_Z) (M_W+M_Z) \\&\quad +
   M_W \sqrt{M_Z^2-M_W^2} C_{HWB}\bigg) C_{00}\left(0,M_Z^2,0,M_W^2,M_T^2,M_T^2\right)\bigg)-\left(2
   M_W^2+M_Z^2\right)^2 \bigg(\sqrt{2} G_{\mu} \left(2 M_W^2+M_Z^2\right) \\&\quad +6  M_W \sqrt{M_Z^2-M_W^2}
   C_{HWB}\bigg) \bigg((D-8) C_0\left(M_Z^2,0,0,0,0,M_Z^2\right) M_Z^2+4 B_0\left(0,0,M_Z^2\right)+(D-6)
   B_0\left(M_Z^2,0,0\right) \\&\quad -2 (D-2) C_{00}\left(0,M_Z^2,0,M_Z^2,0,0\right)\bigg) \Bigg) \numberthis
\end{align*}

\bibliography{zpaper}

\begin{thebibliography}{61}
\expandafter\ifx\csname natexlab\endcsname\relax\def\natexlab#1{#1}\fi
\expandafter\ifx\csname bibnamefont\endcsname\relax
  \def\bibnamefont#1{#1}\fi
\expandafter\ifx\csname bibfnamefont\endcsname\relax
  \def\bibfnamefont#1{#1}\fi
\expandafter\ifx\csname citenamefont\endcsname\relax
  \def\citenamefont#1{#1}\fi
\expandafter\ifx\csname url\endcsname\relax
  \def\url#1{\texttt{#1}}\fi
\expandafter\ifx\csname urlprefix\endcsname\relax\def\urlprefix{URL }\fi
\providecommand{\bibinfo}[2]{#2}
\providecommand{\eprint}[2][]{\url{#2}}

\bibitem[{\citenamefont{Giudice et~al.}(2007)\citenamefont{Giudice, Grojean,
  Pomarol, and Rattazzi}}]{Giudice:2007fh}
\bibinfo{author}{\bibfnamefont{G.~F.} \bibnamefont{Giudice}},
  \bibinfo{author}{\bibfnamefont{C.}~\bibnamefont{Grojean}},
  \bibinfo{author}{\bibfnamefont{A.}~\bibnamefont{Pomarol}}, \bibnamefont{and}
  \bibinfo{author}{\bibfnamefont{R.}~\bibnamefont{Rattazzi}},
  \bibinfo{journal}{JHEP} \textbf{\bibinfo{volume}{06}}, \bibinfo{pages}{045}
  (\bibinfo{year}{2007}), \eprint{hep-ph/0703164}.

\bibitem[{\citenamefont{Brivio and Trott}(2017)}]{Brivio:2017vri}
\bibinfo{author}{\bibfnamefont{I.}~\bibnamefont{Brivio}} \bibnamefont{and}
  \bibinfo{author}{\bibfnamefont{M.}~\bibnamefont{Trott}}
  (\bibinfo{year}{2017}), \eprint{1706.08945}.

\bibitem[{\citenamefont{Buchmuller and Wyler}(1986)}]{Buchmuller:1985jz}
\bibinfo{author}{\bibfnamefont{W.}~\bibnamefont{Buchmuller}} \bibnamefont{and}
  \bibinfo{author}{\bibfnamefont{D.}~\bibnamefont{Wyler}},
  \bibinfo{journal}{Nucl. Phys.} \textbf{\bibinfo{volume}{B268}},
  \bibinfo{pages}{621} (\bibinfo{year}{1986}).

\bibitem[{\citenamefont{Grzadkowski et~al.}(2010)\citenamefont{Grzadkowski,
  Iskrzynski, Misiak, and Rosiek}}]{Grzadkowski:2010es}
\bibinfo{author}{\bibfnamefont{B.}~\bibnamefont{Grzadkowski}},
  \bibinfo{author}{\bibfnamefont{M.}~\bibnamefont{Iskrzynski}},
  \bibinfo{author}{\bibfnamefont{M.}~\bibnamefont{Misiak}}, \bibnamefont{and}
  \bibinfo{author}{\bibfnamefont{J.}~\bibnamefont{Rosiek}},
  \bibinfo{journal}{JHEP} \textbf{\bibinfo{volume}{10}}, \bibinfo{pages}{085}
  (\bibinfo{year}{2010}), \eprint{1008.4884}.

\bibitem[{\citenamefont{Dedes et~al.}(2017)\citenamefont{Dedes, Materkowska,
  Paraskevas, Rosiek, and Suxho}}]{Dedes:2017zog}
\bibinfo{author}{\bibfnamefont{A.}~\bibnamefont{Dedes}},
  \bibinfo{author}{\bibfnamefont{W.}~\bibnamefont{Materkowska}},
  \bibinfo{author}{\bibfnamefont{M.}~\bibnamefont{Paraskevas}},
  \bibinfo{author}{\bibfnamefont{J.}~\bibnamefont{Rosiek}}, \bibnamefont{and}
  \bibinfo{author}{\bibfnamefont{K.}~\bibnamefont{Suxho}},
  \bibinfo{journal}{JHEP} \textbf{\bibinfo{volume}{06}}, \bibinfo{pages}{143}
  (\bibinfo{year}{2017}), \eprint{1704.03888}.

\bibitem[{\citenamefont{Falkowski}(2016)}]{Falkowski:2015fla}
\bibinfo{author}{\bibfnamefont{A.}~\bibnamefont{Falkowski}},
  \bibinfo{journal}{Pramana} \textbf{\bibinfo{volume}{87}}, \bibinfo{pages}{39}
  (\bibinfo{year}{2016}), \eprint{1505.00046}.

\bibitem[{\citenamefont{Contino et~al.}(2014)\citenamefont{Contino, Ghezzi,
  Grojean, Mühlleitner, and Spira}}]{Contino:2014aaa}
\bibinfo{author}{\bibfnamefont{R.}~\bibnamefont{Contino}},
  \bibinfo{author}{\bibfnamefont{M.}~\bibnamefont{Ghezzi}},
  \bibinfo{author}{\bibfnamefont{C.}~\bibnamefont{Grojean}},
  \bibinfo{author}{\bibfnamefont{M.}~\bibnamefont{Mühlleitner}},
  \bibnamefont{and} \bibinfo{author}{\bibfnamefont{M.}~\bibnamefont{Spira}},
  \bibinfo{journal}{Comput. Phys. Commun.} \textbf{\bibinfo{volume}{185}},
  \bibinfo{pages}{3412} (\bibinfo{year}{2014}), \eprint{1403.3381}.

\bibitem[{\citenamefont{Butter et~al.}(2016)\citenamefont{Butter, Éboli,
  Gonzalez-Fraile, Gonzalez-Garcia, Plehn, and Rauch}}]{Butter:2016cvz}
\bibinfo{author}{\bibfnamefont{A.}~\bibnamefont{Butter}},
  \bibinfo{author}{\bibfnamefont{O.~J.~P.} \bibnamefont{Éboli}},
  \bibinfo{author}{\bibfnamefont{J.}~\bibnamefont{Gonzalez-Fraile}},
  \bibinfo{author}{\bibfnamefont{M.~C.} \bibnamefont{Gonzalez-Garcia}},
  \bibinfo{author}{\bibfnamefont{T.}~\bibnamefont{Plehn}}, \bibnamefont{and}
  \bibinfo{author}{\bibfnamefont{M.}~\bibnamefont{Rauch}},
  \bibinfo{journal}{JHEP} \textbf{\bibinfo{volume}{07}}, \bibinfo{pages}{152}
  (\bibinfo{year}{2016}), \eprint{1604.03105}.

\bibitem[{\citenamefont{Ellis et~al.}(2018)\citenamefont{Ellis, Murphy, Sanz,
  and You}}]{Ellis:2018gqa}
\bibinfo{author}{\bibfnamefont{J.}~\bibnamefont{Ellis}},
  \bibinfo{author}{\bibfnamefont{C.~W.} \bibnamefont{Murphy}},
  \bibinfo{author}{\bibfnamefont{V.}~\bibnamefont{Sanz}}, \bibnamefont{and}
  \bibinfo{author}{\bibfnamefont{T.}~\bibnamefont{You}},
  \bibinfo{journal}{JHEP} \textbf{\bibinfo{volume}{06}}, \bibinfo{pages}{146}
  (\bibinfo{year}{2018}), \eprint{1803.03252}.

\bibitem[{\citenamefont{Berthier et~al.}(2016)\citenamefont{Berthier, Bjørn,
  and Trott}}]{Berthier:2016tkq}
\bibinfo{author}{\bibfnamefont{L.}~\bibnamefont{Berthier}},
  \bibinfo{author}{\bibfnamefont{M.}~\bibnamefont{Bjørn}}, \bibnamefont{and}
  \bibinfo{author}{\bibfnamefont{M.}~\bibnamefont{Trott}},
  \bibinfo{journal}{JHEP} \textbf{\bibinfo{volume}{09}}, \bibinfo{pages}{157}
  (\bibinfo{year}{2016}), \eprint{1606.06693}.

\bibitem[{\citenamefont{Falkowski et~al.}(2017)\citenamefont{Falkowski,
  González-Alonso, and Mimouni}}]{Falkowski:2017pss}
\bibinfo{author}{\bibfnamefont{A.}~\bibnamefont{Falkowski}},
  \bibinfo{author}{\bibfnamefont{M.}~\bibnamefont{González-Alonso}},
  \bibnamefont{and} \bibinfo{author}{\bibfnamefont{K.}~\bibnamefont{Mimouni}},
  \bibinfo{journal}{JHEP} \textbf{\bibinfo{volume}{08}}, \bibinfo{pages}{123}
  (\bibinfo{year}{2017}), \eprint{1706.03783}.

\bibitem[{\citenamefont{Hartmann and
  Trott}(2015{\natexlab{a}})}]{Hartmann:2015aia}
\bibinfo{author}{\bibfnamefont{C.}~\bibnamefont{Hartmann}} \bibnamefont{and}
  \bibinfo{author}{\bibfnamefont{M.}~\bibnamefont{Trott}},
  \bibinfo{journal}{Phys. Rev. Lett.} \textbf{\bibinfo{volume}{115}},
  \bibinfo{pages}{191801} (\bibinfo{year}{2015}{\natexlab{a}}),
  \eprint{1507.03568}.

\bibitem[{\citenamefont{Hartmann and
  Trott}(2015{\natexlab{b}})}]{Hartmann:2015oia}
\bibinfo{author}{\bibfnamefont{C.}~\bibnamefont{Hartmann}} \bibnamefont{and}
  \bibinfo{author}{\bibfnamefont{M.}~\bibnamefont{Trott}},
  \bibinfo{journal}{JHEP} \textbf{\bibinfo{volume}{07}}, \bibinfo{pages}{151}
  (\bibinfo{year}{2015}{\natexlab{b}}), \eprint{1505.02646}.

\bibitem[{\citenamefont{Dedes et~al.}(2018)\citenamefont{Dedes, Paraskevas,
  Rosiek, Suxho, and Trifyllis}}]{Dedes:2018seb}
\bibinfo{author}{\bibfnamefont{A.}~\bibnamefont{Dedes}},
  \bibinfo{author}{\bibfnamefont{M.}~\bibnamefont{Paraskevas}},
  \bibinfo{author}{\bibfnamefont{J.}~\bibnamefont{Rosiek}},
  \bibinfo{author}{\bibfnamefont{K.}~\bibnamefont{Suxho}}, \bibnamefont{and}
  \bibinfo{author}{\bibfnamefont{L.}~\bibnamefont{Trifyllis}}
  (\bibinfo{year}{2018}), \eprint{1805.00302}.

\bibitem[{\citenamefont{Gauld et~al.}(2016{\natexlab{a}})\citenamefont{Gauld,
  Pecjak, and Scott}}]{Gauld:2016kuu}
\bibinfo{author}{\bibfnamefont{R.}~\bibnamefont{Gauld}},
  \bibinfo{author}{\bibfnamefont{B.~D.} \bibnamefont{Pecjak}},
  \bibnamefont{and} \bibinfo{author}{\bibfnamefont{D.~J.} \bibnamefont{Scott}},
  \bibinfo{journal}{Phys. Rev.} \textbf{\bibinfo{volume}{D94}},
  \bibinfo{pages}{074045} (\bibinfo{year}{2016}{\natexlab{a}}),
  \eprint{1607.06354}.

\bibitem[{\citenamefont{Gauld et~al.}(2016{\natexlab{b}})\citenamefont{Gauld,
  Pecjak, and Scott}}]{Gauld:2015lmb}
\bibinfo{author}{\bibfnamefont{R.}~\bibnamefont{Gauld}},
  \bibinfo{author}{\bibfnamefont{B.~D.} \bibnamefont{Pecjak}},
  \bibnamefont{and} \bibinfo{author}{\bibfnamefont{D.~J.} \bibnamefont{Scott}},
  \bibinfo{journal}{JHEP} \textbf{\bibinfo{volume}{05}}, \bibinfo{pages}{080}
  (\bibinfo{year}{2016}{\natexlab{b}}), \eprint{1512.02508}.

\bibitem[{\citenamefont{Dawson and
  Giardino}(2018{\natexlab{a}})}]{Dawson:2018pyl}
\bibinfo{author}{\bibfnamefont{S.}~\bibnamefont{Dawson}} \bibnamefont{and}
  \bibinfo{author}{\bibfnamefont{P.~P.} \bibnamefont{Giardino}},
  \bibinfo{journal}{Phys. Rev.} \textbf{\bibinfo{volume}{D97}},
  \bibinfo{pages}{093003} (\bibinfo{year}{2018}{\natexlab{a}}),
  \eprint{1801.01136}.

\bibitem[{\citenamefont{Dawson and
  Giardino}(2018{\natexlab{b}})}]{Dawson:2018liq}
\bibinfo{author}{\bibfnamefont{S.}~\bibnamefont{Dawson}} \bibnamefont{and}
  \bibinfo{author}{\bibfnamefont{P.~P.} \bibnamefont{Giardino}}
  (\bibinfo{year}{2018}{\natexlab{b}}), \eprint{1807.11504}.

\bibitem[{\citenamefont{Hartmann et~al.}(2017)\citenamefont{Hartmann, Shepherd,
  and Trott}}]{Hartmann:2016pil}
\bibinfo{author}{\bibfnamefont{C.}~\bibnamefont{Hartmann}},
  \bibinfo{author}{\bibfnamefont{W.}~\bibnamefont{Shepherd}}, \bibnamefont{and}
  \bibinfo{author}{\bibfnamefont{M.}~\bibnamefont{Trott}},
  \bibinfo{journal}{JHEP} \textbf{\bibinfo{volume}{03}}, \bibinfo{pages}{060}
  (\bibinfo{year}{2017}), \eprint{1611.09879}.

\bibitem[{\citenamefont{Degrande et~al.}(2012)\citenamefont{Degrande, Gerard,
  Grojean, Maltoni, and Servant}}]{Degrande:2012gr}
\bibinfo{author}{\bibfnamefont{C.}~\bibnamefont{Degrande}},
  \bibinfo{author}{\bibfnamefont{J.~M.} \bibnamefont{Gerard}},
  \bibinfo{author}{\bibfnamefont{C.}~\bibnamefont{Grojean}},
  \bibinfo{author}{\bibfnamefont{F.}~\bibnamefont{Maltoni}}, \bibnamefont{and}
  \bibinfo{author}{\bibfnamefont{G.}~\bibnamefont{Servant}},
  \bibinfo{journal}{JHEP} \textbf{\bibinfo{volume}{07}}, \bibinfo{pages}{036}
  (\bibinfo{year}{2012}), \bibinfo{note}{[Erratum: JHEP03,032(2013)]},
  \eprint{1205.1065}.

\bibitem[{\citenamefont{Vryonidou and Zhang}(2018)}]{Vryonidou:2018eyv}
\bibinfo{author}{\bibfnamefont{E.}~\bibnamefont{Vryonidou}} \bibnamefont{and}
  \bibinfo{author}{\bibfnamefont{C.}~\bibnamefont{Zhang}}
  (\bibinfo{year}{2018}), \eprint{1804.09766}.

\bibitem[{\citenamefont{Azatov et~al.}(2017{\natexlab{a}})\citenamefont{Azatov,
  Contino, Machado, and Riva}}]{Azatov:2016sqh}
\bibinfo{author}{\bibfnamefont{A.}~\bibnamefont{Azatov}},
  \bibinfo{author}{\bibfnamefont{R.}~\bibnamefont{Contino}},
  \bibinfo{author}{\bibfnamefont{C.~S.} \bibnamefont{Machado}},
  \bibnamefont{and} \bibinfo{author}{\bibfnamefont{F.}~\bibnamefont{Riva}},
  \bibinfo{journal}{Phys. Rev.} \textbf{\bibinfo{volume}{D95}},
  \bibinfo{pages}{065014} (\bibinfo{year}{2017}{\natexlab{a}}),
  \eprint{1607.05236}.

\bibitem[{\citenamefont{Baglio et~al.}(2017)\citenamefont{Baglio, Dawson, and
  Lewis}}]{Baglio:2017bfe}
\bibinfo{author}{\bibfnamefont{J.}~\bibnamefont{Baglio}},
  \bibinfo{author}{\bibfnamefont{S.}~\bibnamefont{Dawson}}, \bibnamefont{and}
  \bibinfo{author}{\bibfnamefont{I.~M.} \bibnamefont{Lewis}},
  \bibinfo{journal}{Phys. Rev.} \textbf{\bibinfo{volume}{D96}},
  \bibinfo{pages}{073003} (\bibinfo{year}{2017}), \eprint{1708.03332}.

\bibitem[{\citenamefont{Azatov et~al.}(2017{\natexlab{b}})\citenamefont{Azatov,
  Elias-Miro, Reyimuaji, and Venturini}}]{Azatov:2017kzw}
\bibinfo{author}{\bibfnamefont{A.}~\bibnamefont{Azatov}},
  \bibinfo{author}{\bibfnamefont{J.}~\bibnamefont{Elias-Miro}},
  \bibinfo{author}{\bibfnamefont{Y.}~\bibnamefont{Reyimuaji}},
  \bibnamefont{and}
  \bibinfo{author}{\bibfnamefont{E.}~\bibnamefont{Venturini}},
  \bibinfo{journal}{JHEP} \textbf{\bibinfo{volume}{10}}, \bibinfo{pages}{027}
  (\bibinfo{year}{2017}{\natexlab{b}}), \eprint{1707.08060}.

\bibitem[{\citenamefont{Panico et~al.}(2018)\citenamefont{Panico, Riva, and
  Wulzer}}]{Panico:2017frx}
\bibinfo{author}{\bibfnamefont{G.}~\bibnamefont{Panico}},
  \bibinfo{author}{\bibfnamefont{F.}~\bibnamefont{Riva}}, \bibnamefont{and}
  \bibinfo{author}{\bibfnamefont{A.}~\bibnamefont{Wulzer}},
  \bibinfo{journal}{Phys. Lett.} \textbf{\bibinfo{volume}{B776}},
  \bibinfo{pages}{473} (\bibinfo{year}{2018}), \eprint{1708.07823}.

\bibitem[{\citenamefont{Dawson and Valencia}(1995)}]{Dawson:1994fa}
\bibinfo{author}{\bibfnamefont{S.}~\bibnamefont{Dawson}} \bibnamefont{and}
  \bibinfo{author}{\bibfnamefont{G.}~\bibnamefont{Valencia}},
  \bibinfo{journal}{Nucl. Phys.} \textbf{\bibinfo{volume}{B439}},
  \bibinfo{pages}{3} (\bibinfo{year}{1995}), \eprint{hep-ph/9410364}.

\bibitem[{\citenamefont{Zhang}(2017)}]{Zhang:2016zsp}
\bibinfo{author}{\bibfnamefont{Z.}~\bibnamefont{Zhang}},
  \bibinfo{journal}{Phys. Rev. Lett.} \textbf{\bibinfo{volume}{118}},
  \bibinfo{pages}{011803} (\bibinfo{year}{2017}), \eprint{1610.01618}.

\bibitem[{\citenamefont{Alves et~al.}(2018)\citenamefont{Alves, Rosa-Agostinho,
  Éboli, and Gonzalez-Garcia}}]{Alves:2018nof}
\bibinfo{author}{\bibfnamefont{A.}~\bibnamefont{Alves}},
  \bibinfo{author}{\bibfnamefont{N.}~\bibnamefont{Rosa-Agostinho}},
  \bibinfo{author}{\bibfnamefont{O.~J.~P.} \bibnamefont{Éboli}},
  \bibnamefont{and} \bibinfo{author}{\bibfnamefont{M.~C.}
  \bibnamefont{Gonzalez-Garcia}}, \bibinfo{journal}{Phys. Rev.}
  \textbf{\bibinfo{volume}{D98}}, \bibinfo{pages}{013006}
  (\bibinfo{year}{2018}), \eprint{1805.11108}.

\bibitem[{\citenamefont{Hagiwara et~al.}(1992)\citenamefont{Hagiwara, Ishihara,
  Szalapski, and Zeppenfeld}}]{Hagiwara:1992eh}
\bibinfo{author}{\bibfnamefont{K.}~\bibnamefont{Hagiwara}},
  \bibinfo{author}{\bibfnamefont{S.}~\bibnamefont{Ishihara}},
  \bibinfo{author}{\bibfnamefont{R.}~\bibnamefont{Szalapski}},
  \bibnamefont{and}
  \bibinfo{author}{\bibfnamefont{D.}~\bibnamefont{Zeppenfeld}},
  \bibinfo{journal}{Phys. Lett.} \textbf{\bibinfo{volume}{B283}},
  \bibinfo{pages}{353} (\bibinfo{year}{1992}).

\bibitem[{\citenamefont{Hagiwara et~al.}(1993)\citenamefont{Hagiwara, Ishihara,
  Szalapski, and Zeppenfeld}}]{Hagiwara:1993ck}
\bibinfo{author}{\bibfnamefont{K.}~\bibnamefont{Hagiwara}},
  \bibinfo{author}{\bibfnamefont{S.}~\bibnamefont{Ishihara}},
  \bibinfo{author}{\bibfnamefont{R.}~\bibnamefont{Szalapski}},
  \bibnamefont{and}
  \bibinfo{author}{\bibfnamefont{D.}~\bibnamefont{Zeppenfeld}},
  \bibinfo{journal}{Phys. Rev.} \textbf{\bibinfo{volume}{D48}},
  \bibinfo{pages}{2182} (\bibinfo{year}{1993}).

\bibitem[{\citenamefont{Alam et~al.}(1998)\citenamefont{Alam, Dawson, and
  Szalapski}}]{Alam:1997nk}
\bibinfo{author}{\bibfnamefont{S.}~\bibnamefont{Alam}},
  \bibinfo{author}{\bibfnamefont{S.}~\bibnamefont{Dawson}}, \bibnamefont{and}
  \bibinfo{author}{\bibfnamefont{R.}~\bibnamefont{Szalapski}},
  \bibinfo{journal}{Phys. Rev.} \textbf{\bibinfo{volume}{D57}},
  \bibinfo{pages}{1577} (\bibinfo{year}{1998}), \eprint{hep-ph/9706542}.

\bibitem[{\citenamefont{Mebane et~al.}(2013{\natexlab{a}})\citenamefont{Mebane,
  Greiner, Zhang, and Willenbrock}}]{Mebane:2013cra}
\bibinfo{author}{\bibfnamefont{H.}~\bibnamefont{Mebane}},
  \bibinfo{author}{\bibfnamefont{N.}~\bibnamefont{Greiner}},
  \bibinfo{author}{\bibfnamefont{C.}~\bibnamefont{Zhang}}, \bibnamefont{and}
  \bibinfo{author}{\bibfnamefont{S.}~\bibnamefont{Willenbrock}},
  \bibinfo{journal}{Phys. Lett.} \textbf{\bibinfo{volume}{B724}},
  \bibinfo{pages}{259} (\bibinfo{year}{2013}{\natexlab{a}}),
  \eprint{1304.1789}.

\bibitem[{\citenamefont{Mebane et~al.}(2013{\natexlab{b}})\citenamefont{Mebane,
  Greiner, Zhang, and Willenbrock}}]{Mebane:2013zga}
\bibinfo{author}{\bibfnamefont{H.}~\bibnamefont{Mebane}},
  \bibinfo{author}{\bibfnamefont{N.}~\bibnamefont{Greiner}},
  \bibinfo{author}{\bibfnamefont{C.}~\bibnamefont{Zhang}}, \bibnamefont{and}
  \bibinfo{author}{\bibfnamefont{S.}~\bibnamefont{Willenbrock}},
  \bibinfo{journal}{Phys. Rev.} \textbf{\bibinfo{volume}{D88}},
  \bibinfo{pages}{015028} (\bibinfo{year}{2013}{\natexlab{b}}),
  \eprint{1306.3380}.

\bibitem[{\citenamefont{Gaemers and Gounaris}(1979)}]{Gaemers:1978hg}
\bibinfo{author}{\bibfnamefont{K.~J.~F.} \bibnamefont{Gaemers}}
  \bibnamefont{and} \bibinfo{author}{\bibfnamefont{G.~J.}
  \bibnamefont{Gounaris}}, \bibinfo{journal}{Z. Phys.}
  \textbf{\bibinfo{volume}{C1}}, \bibinfo{pages}{259} (\bibinfo{year}{1979}).

\bibitem[{\citenamefont{Hagiwara et~al.}(1987)\citenamefont{Hagiwara, Peccei,
  Zeppenfeld, and Hikasa}}]{Hagiwara:1986vm}
\bibinfo{author}{\bibfnamefont{K.}~\bibnamefont{Hagiwara}},
  \bibinfo{author}{\bibfnamefont{R.~D.} \bibnamefont{Peccei}},
  \bibinfo{author}{\bibfnamefont{D.}~\bibnamefont{Zeppenfeld}},
  \bibnamefont{and} \bibinfo{author}{\bibfnamefont{K.}~\bibnamefont{Hikasa}},
  \bibinfo{journal}{Nucl. Phys.} \textbf{\bibinfo{volume}{B282}},
  \bibinfo{pages}{253} (\bibinfo{year}{1987}).

\bibitem[{\citenamefont{Alonso et~al.}(2014)\citenamefont{Alonso, Jenkins,
  Manohar, and Trott}}]{Alonso:2013hga}
\bibinfo{author}{\bibfnamefont{R.}~\bibnamefont{Alonso}},
  \bibinfo{author}{\bibfnamefont{E.~E.} \bibnamefont{Jenkins}},
  \bibinfo{author}{\bibfnamefont{A.~V.} \bibnamefont{Manohar}},
  \bibnamefont{and} \bibinfo{author}{\bibfnamefont{M.}~\bibnamefont{Trott}},
  \bibinfo{journal}{JHEP} \textbf{\bibinfo{volume}{04}}, \bibinfo{pages}{159}
  (\bibinfo{year}{2014}), \eprint{1312.2014}.

\bibitem[{\citenamefont{Berthier and Trott}(2015)}]{Berthier:2015oma}
\bibinfo{author}{\bibfnamefont{L.}~\bibnamefont{Berthier}} \bibnamefont{and}
  \bibinfo{author}{\bibfnamefont{M.}~\bibnamefont{Trott}},
  \bibinfo{journal}{JHEP} \textbf{\bibinfo{volume}{05}}, \bibinfo{pages}{024}
  (\bibinfo{year}{2015}), \eprint{1502.02570}.

\bibitem[{\citenamefont{Grojean et~al.}(2013)\citenamefont{Grojean, Jenkins,
  Manohar, and Trott}}]{Grojean:2013kd}
\bibinfo{author}{\bibfnamefont{C.}~\bibnamefont{Grojean}},
  \bibinfo{author}{\bibfnamefont{E.~E.} \bibnamefont{Jenkins}},
  \bibinfo{author}{\bibfnamefont{A.~V.} \bibnamefont{Manohar}},
  \bibnamefont{and} \bibinfo{author}{\bibfnamefont{M.}~\bibnamefont{Trott}},
  \bibinfo{journal}{JHEP} \textbf{\bibinfo{volume}{04}}, \bibinfo{pages}{016}
  (\bibinfo{year}{2013}), \eprint{1301.2588}.

\bibitem[{\citenamefont{Hahn}(2001)}]{Hahn:2000kx}
\bibinfo{author}{\bibfnamefont{T.}~\bibnamefont{Hahn}},
  \bibinfo{journal}{Comput. Phys. Commun.} \textbf{\bibinfo{volume}{140}},
  \bibinfo{pages}{418} (\bibinfo{year}{2001}), \eprint{hep-ph/0012260}.

\bibitem[{\citenamefont{Mertig et~al.}(1991)\citenamefont{Mertig, Bohm, and
  Denner}}]{Mertig:1990an}
\bibinfo{author}{\bibfnamefont{R.}~\bibnamefont{Mertig}},
  \bibinfo{author}{\bibfnamefont{M.}~\bibnamefont{Bohm}}, \bibnamefont{and}
  \bibinfo{author}{\bibfnamefont{A.}~\bibnamefont{Denner}},
  \bibinfo{journal}{Comput. Phys. Commun.} \textbf{\bibinfo{volume}{64}},
  \bibinfo{pages}{345} (\bibinfo{year}{1991}).

\bibitem[{\citenamefont{Shtabovenko et~al.}(2016)\citenamefont{Shtabovenko,
  Mertig, and Orellana}}]{Shtabovenko:2016sxi}
\bibinfo{author}{\bibfnamefont{V.}~\bibnamefont{Shtabovenko}},
  \bibinfo{author}{\bibfnamefont{R.}~\bibnamefont{Mertig}}, \bibnamefont{and}
  \bibinfo{author}{\bibfnamefont{F.}~\bibnamefont{Orellana}},
  \bibinfo{journal}{Comput. Phys. Commun.} \textbf{\bibinfo{volume}{207}},
  \bibinfo{pages}{432} (\bibinfo{year}{2016}), \eprint{1601.01167}.

\bibitem[{\citenamefont{Alloul et~al.}(2014)\citenamefont{Alloul, Christensen,
  Degrande, Duhr, and Fuks}}]{Alloul:2013bka}
\bibinfo{author}{\bibfnamefont{A.}~\bibnamefont{Alloul}},
  \bibinfo{author}{\bibfnamefont{N.~D.} \bibnamefont{Christensen}},
  \bibinfo{author}{\bibfnamefont{C.}~\bibnamefont{Degrande}},
  \bibinfo{author}{\bibfnamefont{C.}~\bibnamefont{Duhr}}, \bibnamefont{and}
  \bibinfo{author}{\bibfnamefont{B.}~\bibnamefont{Fuks}},
  \bibinfo{journal}{Comput. Phys. Commun.} \textbf{\bibinfo{volume}{185}},
  \bibinfo{pages}{2250} (\bibinfo{year}{2014}), \eprint{1310.1921}.

\bibitem[{\citenamefont{Christensen et~al.}(2011)\citenamefont{Christensen,
  de~Aquino, Degrande, Duhr, Fuks, Herquet, Maltoni, and
  Schumann}}]{Christensen:2009jx}
\bibinfo{author}{\bibfnamefont{N.~D.} \bibnamefont{Christensen}},
  \bibinfo{author}{\bibfnamefont{P.}~\bibnamefont{de~Aquino}},
  \bibinfo{author}{\bibfnamefont{C.}~\bibnamefont{Degrande}},
  \bibinfo{author}{\bibfnamefont{C.}~\bibnamefont{Duhr}},
  \bibinfo{author}{\bibfnamefont{B.}~\bibnamefont{Fuks}},
  \bibinfo{author}{\bibfnamefont{M.}~\bibnamefont{Herquet}},
  \bibinfo{author}{\bibfnamefont{F.}~\bibnamefont{Maltoni}}, \bibnamefont{and}
  \bibinfo{author}{\bibfnamefont{S.}~\bibnamefont{Schumann}},
  \bibinfo{journal}{Eur. Phys. J.} \textbf{\bibinfo{volume}{C71}},
  \bibinfo{pages}{1541} (\bibinfo{year}{2011}), \eprint{0906.2474}.

\bibitem[{\citenamefont{Shtabovenko}(2017)}]{Shtabovenko:2016whf}
\bibinfo{author}{\bibfnamefont{V.}~\bibnamefont{Shtabovenko}},
  \bibinfo{journal}{Comput. Phys. Commun.} \textbf{\bibinfo{volume}{218}},
  \bibinfo{pages}{48} (\bibinfo{year}{2017}), \eprint{1611.06793}.

\bibitem[{\citenamefont{Patel}(2015)}]{Patel:2015tea}
\bibinfo{author}{\bibfnamefont{H.~H.} \bibnamefont{Patel}},
  \bibinfo{journal}{Comput. Phys. Commun.} \textbf{\bibinfo{volume}{197}},
  \bibinfo{pages}{276} (\bibinfo{year}{2015}), \eprint{1503.01469}.

\bibitem[{\citenamefont{Ellis et~al.}(1996)\citenamefont{Ellis, Stirling, and
  Webber}}]{Ellis:1991qj}
\bibinfo{author}{\bibfnamefont{R.~K.} \bibnamefont{Ellis}},
  \bibinfo{author}{\bibfnamefont{W.~J.} \bibnamefont{Stirling}},
  \bibnamefont{and} \bibinfo{author}{\bibfnamefont{B.~R.}
  \bibnamefont{Webber}}, \bibinfo{journal}{Camb. Monogr. Part. Phys. Nucl.
  Phys. Cosmol.} \textbf{\bibinfo{volume}{8}}, \bibinfo{pages}{1}
  (\bibinfo{year}{1996}).

\bibitem[{\citenamefont{Kniehl}(1991)}]{Kniehl:1991xe}
\bibinfo{author}{\bibfnamefont{B.~A.} \bibnamefont{Kniehl}},
  \bibinfo{journal}{Nucl. Phys.} \textbf{\bibinfo{volume}{B357}},
  \bibinfo{pages}{439} (\bibinfo{year}{1991}).

\bibitem[{\citenamefont{Schwartz}(2014)}]{Schwartz:2013pla}
\bibinfo{author}{\bibfnamefont{M.~D.} \bibnamefont{Schwartz}},
  \emph{\bibinfo{title}{{Quantum Field Theory and the Standard Model}}}
  (\bibinfo{publisher}{Cambridge University Press}, \bibinfo{year}{2014}), ISBN
  \bibinfo{isbn}{1107034736, 9781107034730},
  \urlprefix\url{http://www.cambridge.org/us/academic/subjects/physics/theoretical-physics-and-mathematical-physics/quantum-field-theory-and-standard-model}.

\bibitem[{\citenamefont{Schael et~al.}(2006)}]{ALEPH:2005ab}
\bibinfo{author}{\bibfnamefont{S.}~\bibnamefont{Schael}} \bibnamefont{et~al.}
  (\bibinfo{collaboration}{SLD Electroweak Group, DELPHI, ALEPH, SLD, SLD Heavy
  Flavour Group, OPAL, LEP Electroweak Working Group, L3}),
  \bibinfo{journal}{Phys. Rept.} \textbf{\bibinfo{volume}{427}},
  \bibinfo{pages}{257} (\bibinfo{year}{2006}), \eprint{hep-ex/0509008}.

\bibitem[{\citenamefont{Khachatryan et~al.}(2016)}]{Khachatryan:2015sga}
\bibinfo{author}{\bibfnamefont{V.}~\bibnamefont{Khachatryan}}
  \bibnamefont{et~al.} (\bibinfo{collaboration}{CMS}), \bibinfo{journal}{Eur.
  Phys. J.} \textbf{\bibinfo{volume}{C76}}, \bibinfo{pages}{401}
  (\bibinfo{year}{2016}), \eprint{1507.03268}.

\bibitem[{\citenamefont{Aad et~al.}(2016{\natexlab{a}})}]{Aad:2016wpd}
\bibinfo{author}{\bibfnamefont{G.}~\bibnamefont{Aad}} \bibnamefont{et~al.}
  (\bibinfo{collaboration}{ATLAS}), \bibinfo{journal}{JHEP}
  \textbf{\bibinfo{volume}{09}}, \bibinfo{pages}{029}
  (\bibinfo{year}{2016}{\natexlab{a}}), \eprint{1603.01702}.

\bibitem[{\citenamefont{Aad et~al.}(2016{\natexlab{b}})}]{Aad:2016ett}
\bibinfo{author}{\bibfnamefont{G.}~\bibnamefont{Aad}} \bibnamefont{et~al.}
  (\bibinfo{collaboration}{ATLAS}), \bibinfo{journal}{Phys. Rev.}
  \textbf{\bibinfo{volume}{D93}}, \bibinfo{pages}{092004}
  (\bibinfo{year}{2016}{\natexlab{b}}), \eprint{1603.02151}.

\bibitem[{\citenamefont{Khachatryan et~al.}(2017)}]{Khachatryan:2016poo}
\bibinfo{author}{\bibfnamefont{V.}~\bibnamefont{Khachatryan}}
  \bibnamefont{et~al.} (\bibinfo{collaboration}{CMS}), \bibinfo{journal}{Eur.
  Phys. J.} \textbf{\bibinfo{volume}{C77}}, \bibinfo{pages}{236}
  (\bibinfo{year}{2017}), \eprint{1609.05721}.

\bibitem[{\citenamefont{Aad et~al.}(2016{\natexlab{c}})}]{Khachatryan:2016vau}
\bibinfo{author}{\bibfnamefont{G.}~\bibnamefont{Aad}} \bibnamefont{et~al.}
  (\bibinfo{collaboration}{ATLAS, CMS}), \bibinfo{journal}{JHEP}
  \textbf{\bibinfo{volume}{08}}, \bibinfo{pages}{045}
  (\bibinfo{year}{2016}{\natexlab{c}}), \eprint{1606.02266}.

\bibitem[{\citenamefont{Aaboud et~al.}(2018)}]{Aaboud:2018xdt}
\bibinfo{author}{\bibfnamefont{M.}~\bibnamefont{Aaboud}} \bibnamefont{et~al.}
  (\bibinfo{collaboration}{ATLAS}) (\bibinfo{year}{2018}), \eprint{1802.04146}.

\bibitem[{\citenamefont{Sirunyan et~al.}(2018)}]{Sirunyan:2018ouh}
\bibinfo{author}{\bibfnamefont{A.~M.} \bibnamefont{Sirunyan}}
  \bibnamefont{et~al.} (\bibinfo{collaboration}{CMS}) (\bibinfo{year}{2018}),
  \eprint{1804.02716}.

\bibitem[{\citenamefont{Peskin and Takeuchi}(1990)}]{Peskin:1990zt}
\bibinfo{author}{\bibfnamefont{M.~E.} \bibnamefont{Peskin}} \bibnamefont{and}
  \bibinfo{author}{\bibfnamefont{T.}~\bibnamefont{Takeuchi}},
  \bibinfo{journal}{Phys. Rev. Lett.} \textbf{\bibinfo{volume}{65}},
  \bibinfo{pages}{964} (\bibinfo{year}{1990}).

\bibitem[{\citenamefont{Peskin and Takeuchi}(1992)}]{Peskin:1991sw}
\bibinfo{author}{\bibfnamefont{M.~E.} \bibnamefont{Peskin}} \bibnamefont{and}
  \bibinfo{author}{\bibfnamefont{T.}~\bibnamefont{Takeuchi}},
  \bibinfo{journal}{Phys. Rev.} \textbf{\bibinfo{volume}{D46}},
  \bibinfo{pages}{381} (\bibinfo{year}{1992}).

\bibitem[{\citenamefont{Haller et~al.}(2018)\citenamefont{Haller, Hoecker,
  Kogler, Mönig, Peiffer, and Stelzer}}]{Haller:2018nnx}
\bibinfo{author}{\bibfnamefont{J.}~\bibnamefont{Haller}},
  \bibinfo{author}{\bibfnamefont{A.}~\bibnamefont{Hoecker}},
  \bibinfo{author}{\bibfnamefont{R.}~\bibnamefont{Kogler}},
  \bibinfo{author}{\bibfnamefont{K.}~\bibnamefont{Mönig}},
  \bibinfo{author}{\bibfnamefont{T.}~\bibnamefont{Peiffer}}, \bibnamefont{and}
  \bibinfo{author}{\bibfnamefont{J.}~\bibnamefont{Stelzer}}
  (\bibinfo{year}{2018}), \eprint{1803.01853}.

\bibitem[{\citenamefont{de~Blas et~al.}(2018)\citenamefont{de~Blas, Criado,
  Perez-Victoria, and Santiago}}]{deBlas:2017xtg}
\bibinfo{author}{\bibfnamefont{J.}~\bibnamefont{de~Blas}},
  \bibinfo{author}{\bibfnamefont{J.~C.} \bibnamefont{Criado}},
  \bibinfo{author}{\bibfnamefont{M.}~\bibnamefont{Perez-Victoria}},
  \bibnamefont{and} \bibinfo{author}{\bibfnamefont{J.}~\bibnamefont{Santiago}},
  \bibinfo{journal}{JHEP} \textbf{\bibinfo{volume}{03}}, \bibinfo{pages}{109}
  (\bibinfo{year}{2018}), \eprint{1711.10391}.

\bibitem[{\citenamefont{Chen et~al.}(2014)\citenamefont{Chen, Dawson, and
  Zhang}}]{Chen:2013kfa}
\bibinfo{author}{\bibfnamefont{C.-Y.} \bibnamefont{Chen}},
  \bibinfo{author}{\bibfnamefont{S.}~\bibnamefont{Dawson}}, \bibnamefont{and}
  \bibinfo{author}{\bibfnamefont{C.}~\bibnamefont{Zhang}},
  \bibinfo{journal}{Phys. Rev.} \textbf{\bibinfo{volume}{D89}},
  \bibinfo{pages}{015016} (\bibinfo{year}{2014}), \eprint{1311.3107}.

\end{thebibliography}
\end{document}